\documentclass{revtex4}

\usepackage{amsthm,amssymb,amsmath}				
\usepackage{graphicx,comment}
\usepackage{float}   
\usepackage{xcolor}

\begin{document}

\title{Confined Klein-Gordon oscillators in Minkowski spacetime and a
pseudo-Minkowski spacetime with a space-like dislocation: PDM
KG-oscillators, isospectrality and invariance}
\author{Omar Mustafa}
\email{omar.mustafa@emu.edu.tr}
\affiliation{Department of Physics, Eastern Mediterranean University, G. Magusa, north
Cyprus, Mersin 10 - Turkey.}

\begin{abstract}
\textbf{Abstract:}\ We revisit the a confined (in a Cornell-type Lorentz
scalar potential) KG-oscillator in Minkowski spacetime with space-like
dislocation background. We show that the effect of space-like dislocation is
to shift the energy levels along the dislocation parameter axis, and
consequently energy levels crossings are unavoidable. We report some
KG-particles in a pseudo-Minkowski spacetime with space-like dislocation
that admit isospectrality and invariance with the confined KG-oscillator in
Minkowski spacetime with space-like dislocation. An alternative PDM setting
for the KG-particles (relativistic particles in general) is introduced. We
discuss the effects of space-like dislocation and PDM settings on the
confined KG-oscillators in Minkowski spacetime with space-like dislocation.
Three confined PDM KG-oscillators are discussed as illustrative examples,
(i) a PDM KG-oscillator from a dimensionless scalar multiplier $g\left(
r\right) =\, exp(2\alpha r^2)\geq0,\, \alpha\,\geq 0$, (ii) a PDM
KG-oscillator from a power law type dimensionless scalar multiplier $g\left(
r\right) =Ar^{\sigma }\geq0$, and (iii) a PDM KG-oscillator in a
Cornnell-type confinement with a dimensionless scalar multiplier $g\left(
r\right) =\exp \left( \xi r\right)\geq0$ .

\textbf{PACS }numbers\textbf{: }03.65.Ge,03.65.Pm,02.40.Gh

\textbf{Keywords:} Klein-Gordon (KG) oscillator, spacetime with space-like
screw dislocation, position-dependent mass KG-particles.
\end{abstract}

\maketitle

\section{Introduction}

The grand unified theories have predicted possible topological defects in spacetime \cite{Kibble 1980,Vilenkin 1981,Vilenkin 1985,Braganca 2020} that have been investigated in many areas of physics. For example, in condensed matter physics \cite{Katanaev 1992}, in gravitation \cite{Puntigam 1997,da Silva 2019,Vitoria 2019} (where the linear defects are due dislocation (torsion) and curvature (disclinations)), in domain wall \cite{Vilenkin 1981,Vilenkin 1985}, in cosmic string \cite{Vilenkin 1983,Linet 1985}, in global monopole \cite{Barriola 1989}, etc. However, in their work on Volterra distortions and cosmic defects, Puntigam and Soleng \cite{Puntigam 1997} have generalized the Volterra distortion to (3+1)-dimensions, using differential geometric and gauge theoretical methods, and introduced the concept of Volterra distorted spacetime. Where, distortions are line-like defects characterized by a delta-function-valued curvature (classified as disclination) and torsion (classified as dislocation) distributions that result in rotational and translational holonomy. Dislocation may be in the form of a spiral-type \cite{da Silva 2019} or a screw-type \cite{Vitoria 2018,Vitoria 2019}. The latter is in point of the current study.

Such topological defects in spacetime have their figure prints on the spectroscopic structure of relativistic and non-relativistic quantum systems. The Dirac oscillator \cite{Moshinsky 1989}, for example, is investigated by Hassanabadi and co-worker \cite{Montigny 2018,Sedaghatnia 2019} in a G\"{o}del-type cosmic string Som-Raychaudhuri spacetime. The Klein-Gordon (KG) oscillator is studied in the G\"{o}del-type spacetime (e.g., \cite{Bruce 1993,Dvoeg 1994,Das 2008,Carvalho 2016,Garcia 2017,Vitoria 2016,Vitoria 2018}), in cosmic string spacetime and Kaluza-Klein theory (e.g., \cite{Vitoria 2018,Ahmed 2020,Ahmed1 2020,Ahmed1 2021,Boumal 2014,Mustafa2 2022}), in Som-Raychaudhuri spacetime \cite{Wang 2015}, in the (2+1)-dimensional G\"{u}rses spacetime (e.g., \cite{Gurses 1994,Ahmed 2019,Ahmed1 2019,Ahmed2 2019,Mustafa1 2022}).

On the other hand, the concept of position-dependent effective mass (PDM) (initiated by Mathews-Lakshmanan oscillator \cite{M-L 1974}) has sparked research interest on PDM in both classical and quantum mechanics \cite{M-L 1974,von Roos,Carinena Ranada Sant 2004,Mustafa 2019,Mustafa 2020,Mustafa arXiv, Mustafa Phys.Scr. 2020,Mustafa Habib 2007,Mustafa Algadhi 2019,Zeinab 2020,Khlevniuk 2018,Mustafa 2015,Dutra Almeida 2000,dos Santos 2021,Nabulsi1 2020,Nabulsi2 2020,Nabulsi3 2021,Quesne 2015,Tiwari 2013,Alimohammadi 2017,Pourali 2021,Ikot 2016,Ghabab 2016}. Such a PDM concept is, in fact, a \textit{metaphoric manifestation of coordinate transformation} \cite{Mustafa 2020,Mustafa arXiv,Mustafa Phys.Scr. 2020,Khlevniuk 2018}. The coordinate transformation, in effect, changes the form of the canonical momentum in classical and the momentum operator in quantum mechanics (e.g., \cite{Mustafa 2020,Mustafa arXiv,Mustafa Algadhi 2019,dos Santos 2021} and related references therein). In classical mechanics, for example, negative the gradient of the potential force field is no longer the time derivative of the canonical momentum $p=m\left(x\right) \dot{x};\,m(x)=m\,g(x)$, but it is rather related to the time derivative of the pseudo-momentum (also called Noether momentum) $\pi\left( x\right) =\sqrt{m\,g\left( x\right) }\dot{x}$ \cite{Mustafa arXiv},
where $m$ denotes the rest mass of the classical/quantum mechanical particle at hand and $g(x)$ is a dimensionless positive valued scalar multiplier. Expressing the canonical momentum as $p=m\left(x\right) \dot{x}$ with $\,m(x)=m\,g(x)$ inevitably suggests the metaphor notion of PDM particles. In quantum mechanics, moreover, the PDM (metaphorically speaking) momentum operator is constructed by Mustafa and Algadhi \cite{Mustafa Algadhi 2019} to read%
\begin{equation}
\mathbf{\hat{p}}\left( \mathbf{r}\right) =-i\left( \mathbf{\nabla -}\frac{%
\mathbf{\nabla }g\left( \mathbf{r}\right) }{4\,g\left( \mathbf{r}\right) }%
\right) \Longleftrightarrow \hat{p}_{j}\left( \mathbf{r}\right)=-i\left(
\partial _{j}-\frac{\partial _{j}g\left( \mathbf{r}\right) }{4\,g\left( 
\mathbf{r}\right) }\right); \, j=1,2,3.  \label{PDM-op}
\end{equation}%
Which, in its most simplistic one-dimensional form, suggests (e.g., \cite{Mustafa 2020,Mustafa Habib 2007} for more details) that the von Roos \cite{von Roos} PDM kinetic energy operator, in $\hbar =1$ units,%
\begin{equation*}
\hat{T}\left( x\right) \Phi(x)=\left(\frac{\hat{p}_x\left( x\right)}{\sqrt{%
2\,m(x)}}\right)^2\Phi(x)=-\frac{1}{2m}g(x)^{-1/4}\, \partial_x
g(x)^{-1/2}\,\partial_x g(x)^{-1/4}\Phi(x),  \label{PDM-kinetic op}
\end{equation*}%
or, in $\hbar =2m=1$ units, the von Roos \cite{von Roos} PDM kinetic energy operator reads %
\begin{equation}
\hat{T}\left( x\right) \Phi(x)=\left(\frac{\hat{p}_x\left( x\right)}{\sqrt{g(x)}}\right)^2\Phi(x)=-g(x)^{-1/4}\, \partial_x g(x)^{-1/2}\partial_x
g(x)^{-1/4}\Phi(x),  \label{PDM-kinetic op}
\end{equation}%
which is known in the literature as Mustafa and Mazharimousavi's ordering of the ambiguity parameters involved in the von Roos \cite{von Roos} PDM kinetic energy operator \cite{Mustafa Habib 2007}. This result clearly indicates that the momentum operator of an effective and metaphoric PDM quantum particle is given by (\ref{PDM-op}). It has also been reported that such PDM quantum particles (as so should be metaphorically called hereinafter) may very well be trapped in their own byproducted force fields (i.e., quasi-free PDM particles is used to describe such a system \cite{Zeinab 2020}). Yet, it has been used to find the PDM creation and annihilation operators for the PDM-Schr\"{o}dinger oscillator \cite{Mustafa 2020}.

Nevertheless, attempts were made to include PDM settings in the Dirac and KG relativistic equations through the assumption that $m\longrightarrow m+S\left( r\right)=m(r) $, where $m$ denotes the rest mass energy, $S\left(r\right)$ is the Lorentz scalar potential (commonly used in heavy quarkonium spectroscopy \cite{Quigg 1979}), and $m(r)$ denotes PDM (e.g., \cite{Mustafa Habib 2008,Mustafa Habib1 2007,Vitoria Bakke 2016}). In the current study, however, we shall not use this assumption but rather argue
that analogous to textbook procedure, where the momentum operator $p_j=-i\partial_j$ for constant mass is used in the relativistic wave equations, so should be the case with the PDM-momentum operator (\ref{PDM-op}) to describe PDM-relativistic quantum particles (e.g., \cite{Mustafa Algadhi 2019,Zeinab 2020,Mustafa 2020,dos Santos 2021}). That is, for PDM particles (relativistic and non-relativistic), the PDM-momentum operator (\ref{PDM-op}) should replace the constant mass textbook momentum operator $p_j=-i\partial_j$. In the current methodical proposal, we use such a PDM assumption and investigate the effects of the gravitational field generated by Minkowski spacetime with space-like screw dislocation ((\ref{screw metric 1}) below) on some confined PDM KG-oscillators.

The organization of this paper is in order. We revisit, in section 2, with a confined (in a Cornnell-type potential \cite{Quigg 1979,Lutfuoglu 2020}) KG-oscillator in Minkowski spacetime with space-like dislocation background. We show that the effect of space-like dislocation is to shift the energy levels along the dislocation parameter axis, and consequently energy levels crossings (i.e., occasional degeneracies) are unavoidable. Energy levels crossings, nevertheless, is a phenomenon responsible for electron transfer in protein, it underlies stability analysis in mechanical engineering, and appears in algebraic geometry (e.g., \cite{Bhattacharya 2006} and references
cited therein). Moreover, clusterings of energy levels are found feasible for $\left\vert \delta \right\vert >>1$, where $\delta $ denotes space-like dislocation parameter. Taking our analysis of section 2 into account, we report, in section 3, some KG-particles in a transformed pseudo-Minkowski spacetime with space-like dislocation (\ref{screw metric 2}), below, that admit isospectrality and invariance with the KG-oscillator in Minkowski spacetime with space-like dislocation (\ref{screw metric 1}), below.
Moreover, we suggest (in section 4) an alternative PDM setting for the KG-particles (relativistic particles in general). Therein, we use the PDM-momentum operator (\ref{PDM-op}), constructed by Mustafa and Algadhi \cite{Mustafa Algadhi 2019}, and discuss the effects of space-like dislocation and PDM settings on the confined KG-oscillators in Minkowski spacetime with space-like dislocation. Three confined PDM KG-oscillators are used/discussed as illustrative examples, (i) a PDM KG-oscillator from a dimensionless scalar multiplier $g\left( r\right) =\, exp(2\alpha r^2)\geq0,\, \alpha\,\geq 0$, (ii) a PDM KG-oscillator from a power law type dimensionless scalar multiplier $g\left( r\right) =Ar^{\sigma }\geq0$, and (iii) a PDM KG-oscillator in a Cornnell-type confinement with a dimensionless scalar multiplier $g\left( r\right) =\exp \left( \xi r\right)\geq0$ . Our concluding remarks are given in section 5.

\section{Confined KG-oscillator in Minkowski spacetime with space-like screw dislocation: revisited}

In this section, we consider Minkowski spacetime with space-like screw dislocation (i.e., a Volterra-type spacetime with space-like dislocation 
\cite{Puntigam 1997,Lima 2017,Bakke 2021,Vitoria 2019} (in $\hbar =c=1$ units) described by the line element%
\begin{equation}
ds^{2}=-dt^{2}+dr^{2}+r^{2}d\varphi ^{2}+\left( dz+\delta d\varphi \right)
^{2},  \label{screw metric 1}
\end{equation}%
where $\delta $ denotes space-like dislocation parameter (i.e., torsion parameter). The covariant and contravariant metric tensors in this case, respectively, read%
\begin{equation}
g_{\mu \nu }=\left( 
\begin{tabular}{cccc}
$-1\smallskip $ & $\,0\,$ & $0$ & $\,0$ \\ 
$0$ & $1\smallskip $ & $0$ & $0$ \\ 
$0$ & $\,0$ & $\,\left( r^{2}+\delta ^{2}\right) \,$ & $\delta $ \\ 
$0$ & $0$ & $\delta $ & $1$%
\end{tabular}%
\right) \Longleftrightarrow g^{\mu \nu }=\left( 
\begin{tabular}{cccc}
$-1\smallskip $ & $\,0\,$ & $0$ & $0$ \\ 
$0$ & $\,1\smallskip $ & $0$ & $0$ \\ 
$0$ & $\,0$ & $\,\frac{1}{r^{2}}\,$ & $-\frac{\delta }{r^{2}}$ \\ 
$0$ & $0$ & $-\frac{\delta }{r^{2}}\,$ & $\,\left( 1+\frac{\delta ^{2}}{r^{2}%
}\right) $%
\end{tabular}%
\right) \text{ };\text{ \ }\det \left( g\right) =-r^{2}.  \label{g-metric1}
\end{equation}%
On the other hand, the KG-equation, with a Lorentz scalar potential $S\left(r\right) $ (i.e., $m\longrightarrow m+S\left( r\right) $) \cite{Mustafa Habib 2008,Mustafa Habib1 2007}, is given by%
\begin{equation}
\frac{1}{\sqrt{-g}}\partial _{\mu }\left( \sqrt{-g}g^{\mu \nu }\partial
_{\nu }\Psi \right) =\left( m+S\left( r\right) \right) ^{2}\Psi .  \label{KG}
\end{equation}%
Moreover, we may now use similar recipe to those used in \cite{Moshinsky 1989,Ahmed 2020,Lutfuoglu 2020,Mirza 2004} and consider%
\begin{equation}
p_{\mu }\longrightarrow p_{\mu }+i\eta \chi _{\mu },  \label{momentum op}
\end{equation}%
where $\chi _{\mu }=\left( 0,r,0,0\right) $. This would, in effect, transform KG-equation (\ref{KG}) into%
\begin{equation}
\frac{1}{\sqrt{-g}}\left( \partial _{\mu }+\eta \chi _{\mu }\right) \left[ 
\sqrt{-g}g^{\mu \nu }\left( \partial _{\nu }-\eta \chi _{\nu }\right) \Psi %
\right] =\left( m+S\left( r\right) \right) ^{2}\Psi ,
\label{KG-oscillator 1}
\end{equation}%
Hereby, one may use $\eta =m\omega \geq 0$ (with $m$ denoting rest mass of the KG-particle) to recover the traditionally used values as in (e.g., \cite{Vitoria 2018} and other related references cited therein). However, we shall use a more general parameter $\eta \geq 0$ . In this case, we avoid eminent confusion and inconsistency between $m^{2}$ (denoting $m^{2}c^{2}=m m c^{2}$, the rest mass multiplied by the rest mass energy, should the KG-equation (\ref{KG-oscillator 1}) be divided by $c^{2}$) on the
R.H.S. and the rest mass of the particle on the L.H.S. of the KG-equation (\ref{KG-oscillator 1}) for $\eta =m\omega $ case. This point is made implicitly clear by Moshinsky and Szczepaniak \cite{Moshinsky 1989} and Mirza and Mohadesi \cite{Mirza 2004} while dealing with the Dirac and KG oscillators, who kept the speed of light $c$ as is. We therefore stick with our assumption and use the spacetime metric tensor elements in (\ref{g-metric1}), to recast (\ref{KG-oscillator 1}) as%
\begin{equation}
\left\{ -\partial _{t}^{2}+\left( \partial _{r}^{2}+\frac{1}{r}\partial
_{r}\right) +\frac{1}{r^{2}}\partial _{\varphi }^{2}+\left( 1+\frac{\delta
^{2}}{r^{2}}\right) \partial _{z}^{2}-\frac{2\delta }{r^{2}}\partial
_{\varphi }\partial _{z}-\eta ^{2}r^{2}-2\eta -\left( m+S\left( r\right)
\right) ^{2}\right\} \Psi =0.  \label{KG-eq0}
\end{equation}%
A substitution in the form of%
\begin{equation}
\Psi \left( t,r,\varphi ,z\right) =\exp \left( i\left[ \ell \varphi
+k_{z}z-Et\right] \right) \psi \left( r\right) =\exp \left( i\left[ \ell
\varphi +k_{z}z-Et\right] \right) \frac{R\left( r\right) }{\sqrt{r}}
\label{Psi 1}
\end{equation}%
would result in%
\begin{equation}
R^{\prime \prime }\left( r\right) +\left[ \lambda -\frac{\left( \tilde{\ell}%
^{2}-1/4\right) }{r^{2}}-\eta ^{2}r^{2}-2mS\left( r\right) -S\left( r\right)
^{2}\right] R\left( r\right) =0,  \label{radial eq 1}
\end{equation}%
where%
\begin{equation}
\lambda =E^{2}-k_{z}^{2}-2\eta -m^{2}\,;\,\,\tilde{\ell}^{2}=\left( \ell
-k_{z}\delta \right) ^{2}.  \label{parameters 1}
\end{equation}%
Notably, the effect of the space-like dislocation is to introduce a shift in the irrational magnetic quantum number $\tilde{\ell}=\pm |\ell -k_{z}\delta |$ of (\ref{parameters 1}), where $\ell =0,\pm 1,\pm 2.\cdots $, is the magnetic quantum number. Moreover, equation (\ref{radial eq 1}) resembles, with $S(r)=0$, the two-dimensional radial Schr\"{o}dinger oscillator (in the units $2m=\hbar =1$) with an effective oscillation frequency $\eta \geq 0$. Consequently and mathematically inherits its textbook eigenvalues%
\begin{equation}
\lambda =2\eta \left( 2n_{r}+\left\vert \tilde{\ell}\right\vert +1\right)
\Longleftrightarrow E^{2}=2\eta \left( 2n_{r}+\left\vert \ell -k_{z}\delta
\right\vert +2\right) +k_{z}^{2}+m^{2}  \label{2D-lambda 1}
\end{equation}%
and radial eigenfunctions%
\begin{equation}
\psi \left( r\right) \sim r^{\left\vert \ell -k_{z}\delta \right\vert }\exp
\left( -\frac{\eta r^{2}}{2}\right) L_{n_{r}}^{\left\vert \ell -k_{z}\delta
\right\vert }\left( \eta r^{2}\right) \Longleftrightarrow \psi \left(
r\right) \sim r^{\left\vert \ell -k_{z}\delta \right\vert }\exp \left( -%
\frac{\eta r^{2}}{2}\right) L_{n_{r}}^{\left\vert \ell -k_{z}\delta
\right\vert }\left( \eta r^{2}\right) ,  \label{radial wave 1}
\end{equation}%
\begin{figure}[!ht]  
\centering
\includegraphics[width=0.3\textwidth]{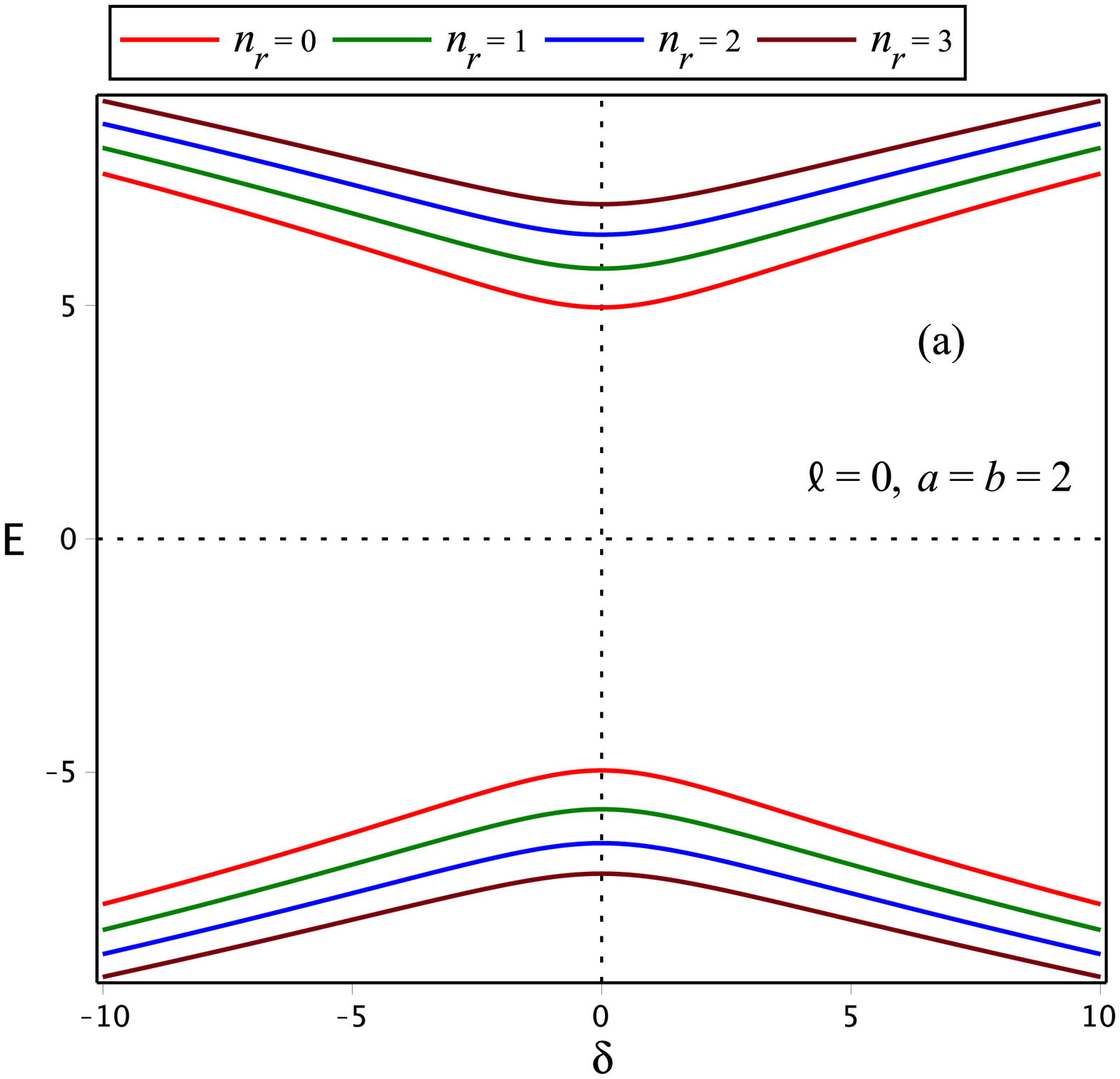}
\includegraphics[width=0.3\textwidth]{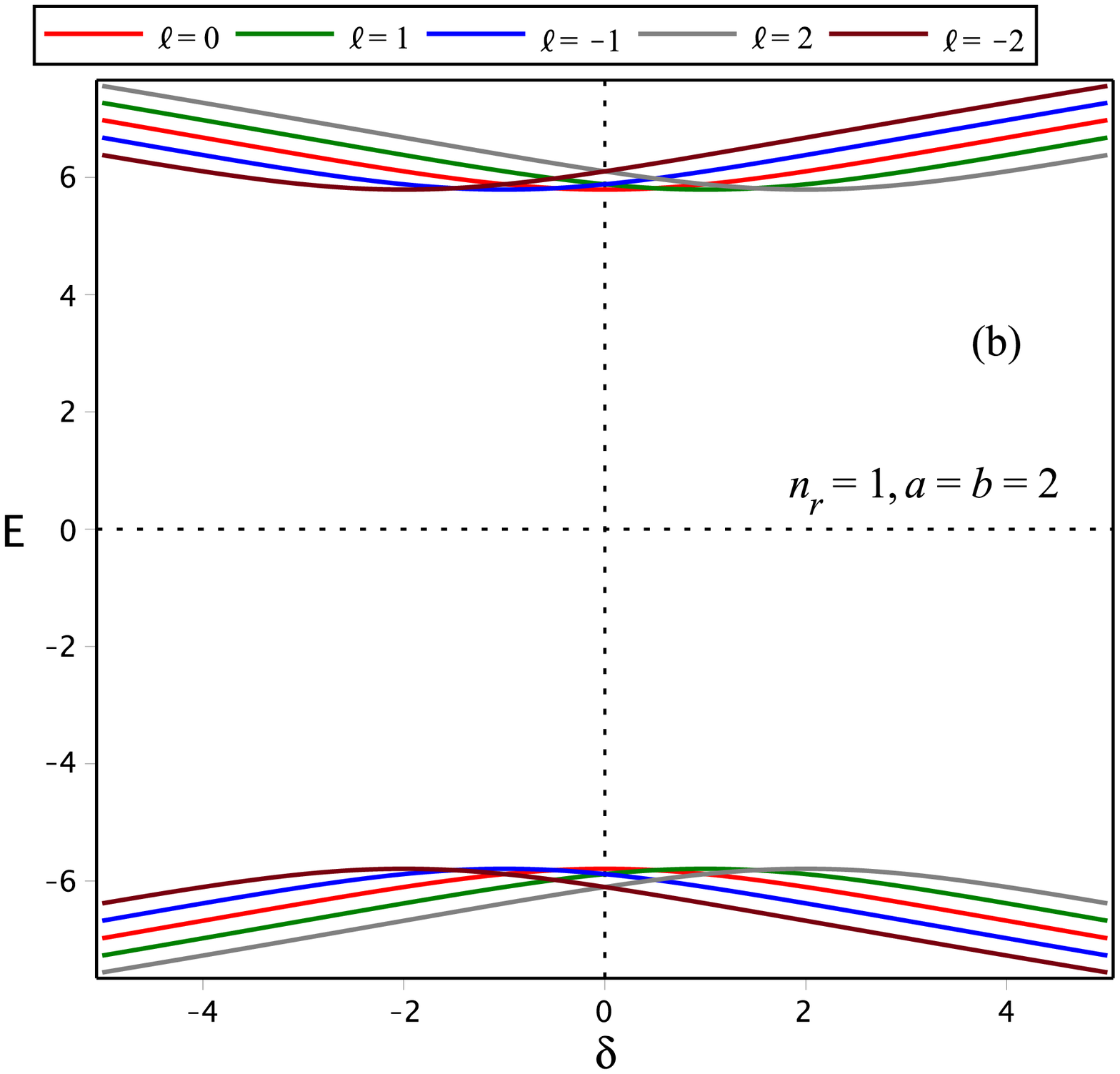}
\includegraphics[width=0.3\textwidth]{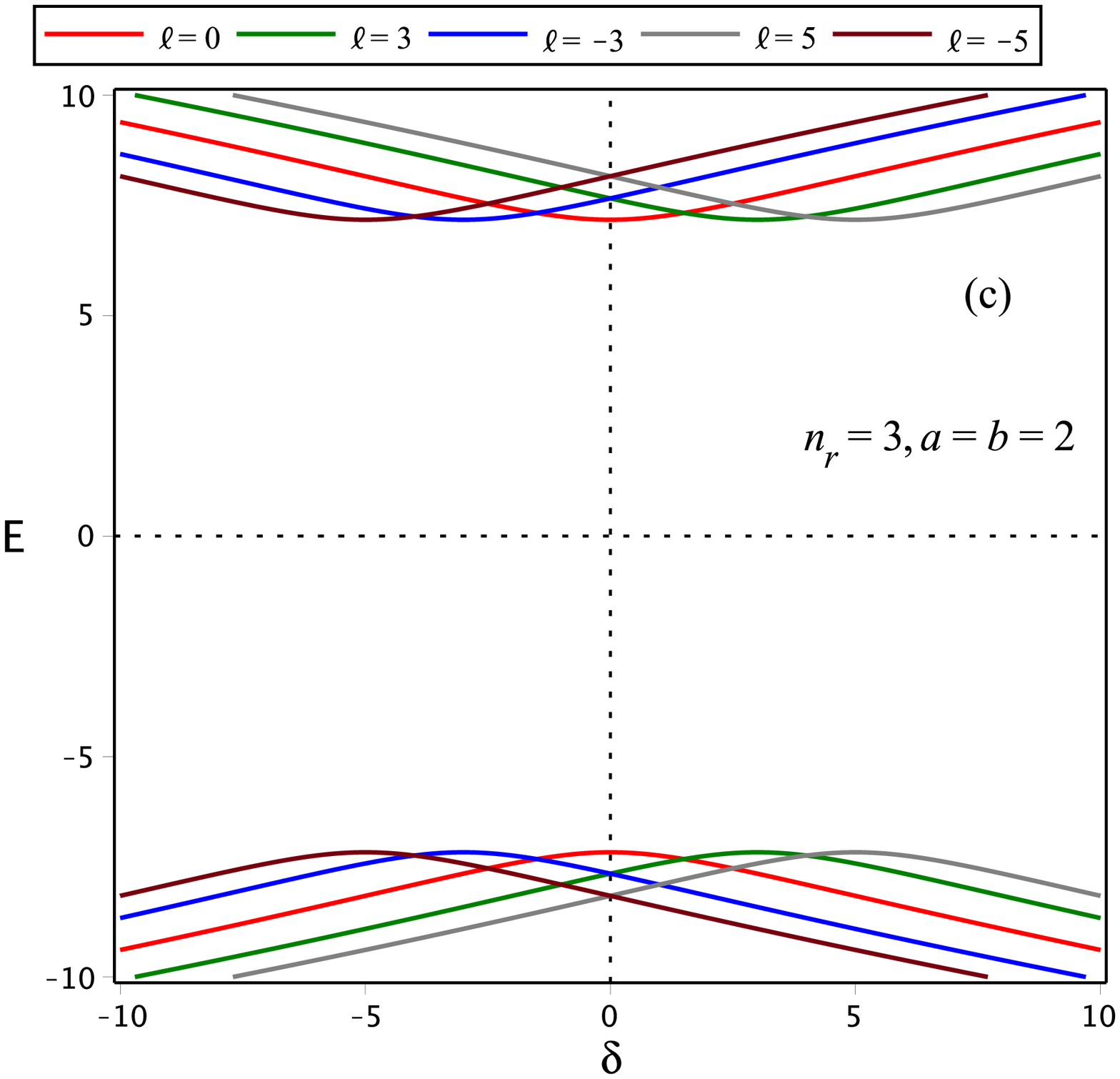}
\caption{\small 
{ We plot the energy levels of (\ref{energy eq 1}) versus the torsion parameter $\delta$, for $m=k_z=\eta=1$, $a=b=2$ and for (a) $\ell=0$, $n_r=0,1,2,3$, (b) $n_r=1$, $\ell=0,\pm1,\pm2$, and (c) $n_r=3$, $\ell=0,\pm3,\pm5$.}}
\label{fig1}
\end{figure}%
where $L_{n_{r}}^{\left\vert \ell -k_{z}\delta \right\vert }\left( \eta r^{2}\right) $ are the the associated Laguerre polynomial. Moreover, one should notice that our results in (\ref{2D-lambda 1}) and (\ref{radial wave 1}) exactly agree with those reported by Carvalho et al \cite{Carvalho 2016} (equation (25) of Carvallho), by Vit\'{o}ria and Bakke \cite{R1} (their result (31) an (30), respectively), and by Medeirosa and de Mello \cite{R2} (with the missed terms added in their (36), (38), (59), (61), i.e., 
$k^{2}\rightarrow k^{2}+M^{2}$, and their $k^{2}\rightarrow k^{2}+M^{2}-\delta _{L}^{2}/4\Delta $ in (49) and (51)) , of course with the proper parametric matching.

Let us now consider the KG-oscillator above be confined in a Cornell type potential%
\begin{equation}
S\left( r\right) =a r+\frac{b}{r}.  \label{Cornell potential}
\end{equation}%
In this case, equation (\ref{radial eq 1}) reads%
\begin{equation}
R^{\prime \prime }\left( r\right) +\left[ \tilde{\lambda}-\frac{\left( 
\tilde{\gamma}^{2}-1/4\right) }{r^{2}}-\tilde{\omega}^{2}r^{2}-2mar-\frac{2mb%
}{r}\right] R\left( r\right) =0,  \label{cornell-1 R(r) eq}
\end{equation}%
where%
\begin{equation}
\tilde{\lambda}=E^{2}-k_{z}^{2}-2\eta -m^{2}-2ab\,;\,\ \tilde{\gamma}%
^{2}=\left( \ell -k_{z}\delta \right) ^{2}+b^{2}\,;\ \tilde{\omega}^{2}=\eta
^{2}+a^{2}.  \label{Cornell eq parameters}
\end{equation}%
Now, $\tilde{\gamma}=\pm \sqrt{\left( \ell -k_{z}\delta \right) ^{2}+b^{2}}$ is the new irrational magnetic quantum number and $\tilde{\omega}=\sqrt{\eta^{2}+a^{2}}\geq 0$ is our new effective oscillation frequency. Equation (\ref{cornell-1 R(r) eq}) admits a solution in the form of%
\begin{equation}
\psi (r)=\frac{R\left( r\right) }{\sqrt{r}}\sim \mathcal{\,}r^{\left\vert 
\tilde{\gamma}\right\vert }\,\exp \left( -\frac{\ \tilde{\omega}%
^{2}r^{2}+2amr}{2\,\tilde{\omega}}\right) \,H_{B}\left( 2\left\vert \tilde{%
\gamma}\right\vert ,\frac{2ma}{\ \tilde{\omega}^{3/2}},\frac{a^{2}m^{2}+%
\tilde{\lambda}\,\ \tilde{\omega}^{2}}{\ \tilde{\omega}^{3}},\frac{4mb}{%
\sqrt{\ \tilde{\omega}}},\sqrt{\ \tilde{\omega}}r\right) ,
\label{S(r)-solution}
\end{equation}%
where $H_{B}\left( \alpha ,\beta ,\gamma ,\delta ,r\right) $ is the biconfluent Heun function that is truncated into polynomial of degree $n\geq 0$ by the condition that $\gamma =2\left( n+1\right) +\alpha $ to secure finiteness and square integrability of the solution. However, the truncation condition would provide a quantization recipe but does not make $n$ a valid quantum number. In this case, if we set $n=2n_{r}\geq 0$, where $n_{r}=0,1,2,\cdots $ is the radial quantum number then the condition $\gamma
=2\left( 2n_{r}+1\right) +\alpha $ would satisfy Ronveaux's condition \cite{Ron 1995,Neto 2020} and implies%
\begin{equation}
\frac{a^{2}m^{2}+\tilde{\lambda}\,\tilde{\omega}^{2}}{\tilde{\omega}^{3}}%
=2\left( 2n_{r}+\left\vert \tilde{\gamma}\right\vert +1\right)
\Longleftrightarrow \tilde{\lambda}=2\tilde{\omega}\left( 2n_{r}+\left\vert 
\tilde{\gamma}\right\vert +1\right) -\frac{m^{2}a^{2}}{\tilde{\omega}^{2}}.
\label{lambda-S(r)}
\end{equation}%
Hence, we get the relation for the energy eigenvalues as%
\begin{equation}
E^{2}=2\left( \sqrt{\eta ^{2}+a^{2}}\right) \left( 2n_{r}+\left\vert \sqrt{%
\left( \ell -k_{z}\delta \right) ^{2}+b^{2}}\right\vert +1\right) -\frac{%
m^{2}a^{2}}{\eta ^{2}+a^{2}}+2\eta +k_{z}^{2}+m^{2}+2ab.  \label{energy eq 1}
\end{equation}%
\begin{figure}[!ht]  
\centering
\includegraphics[width=0.3\textwidth]{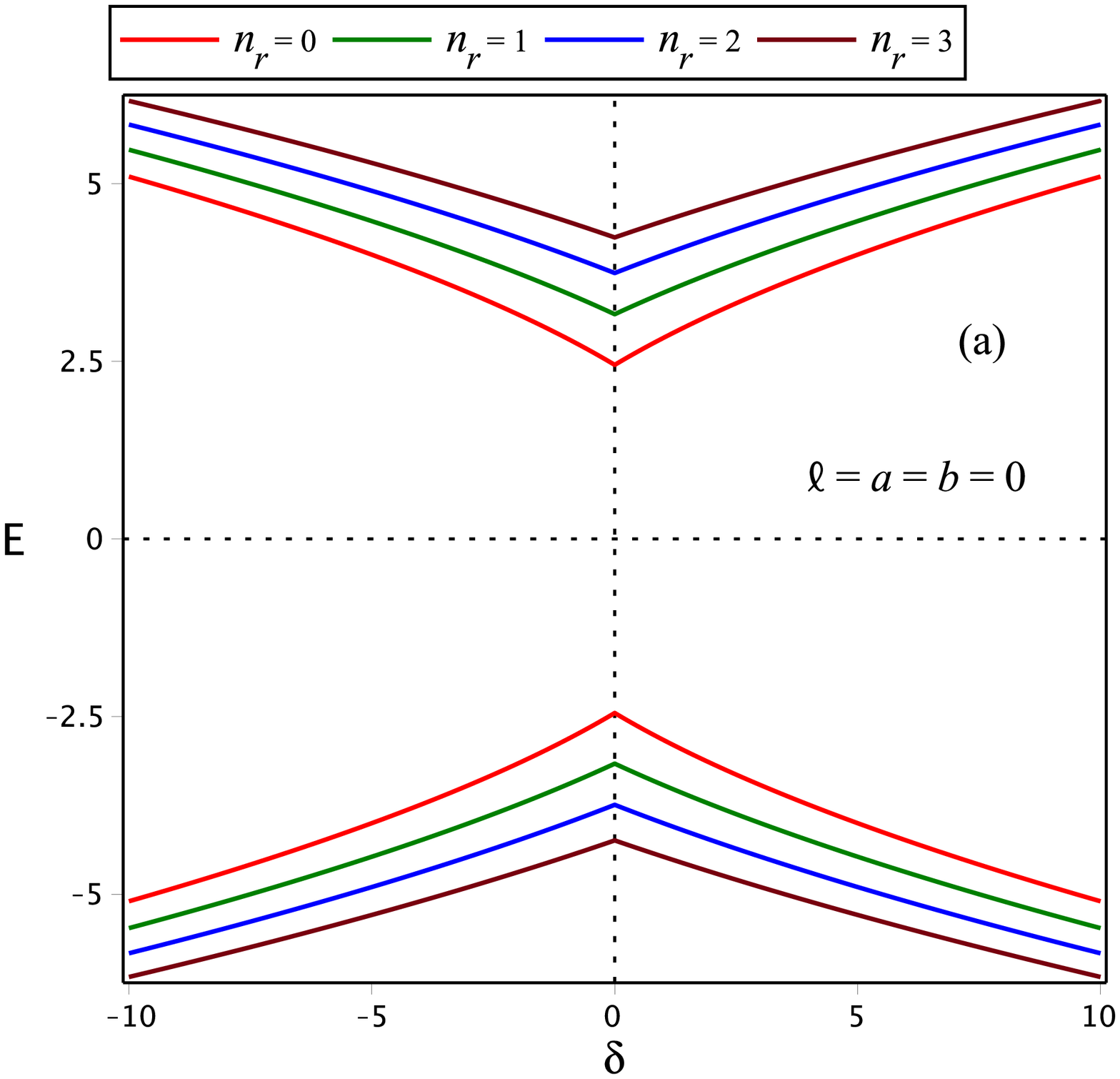}
\includegraphics[width=0.3\textwidth]{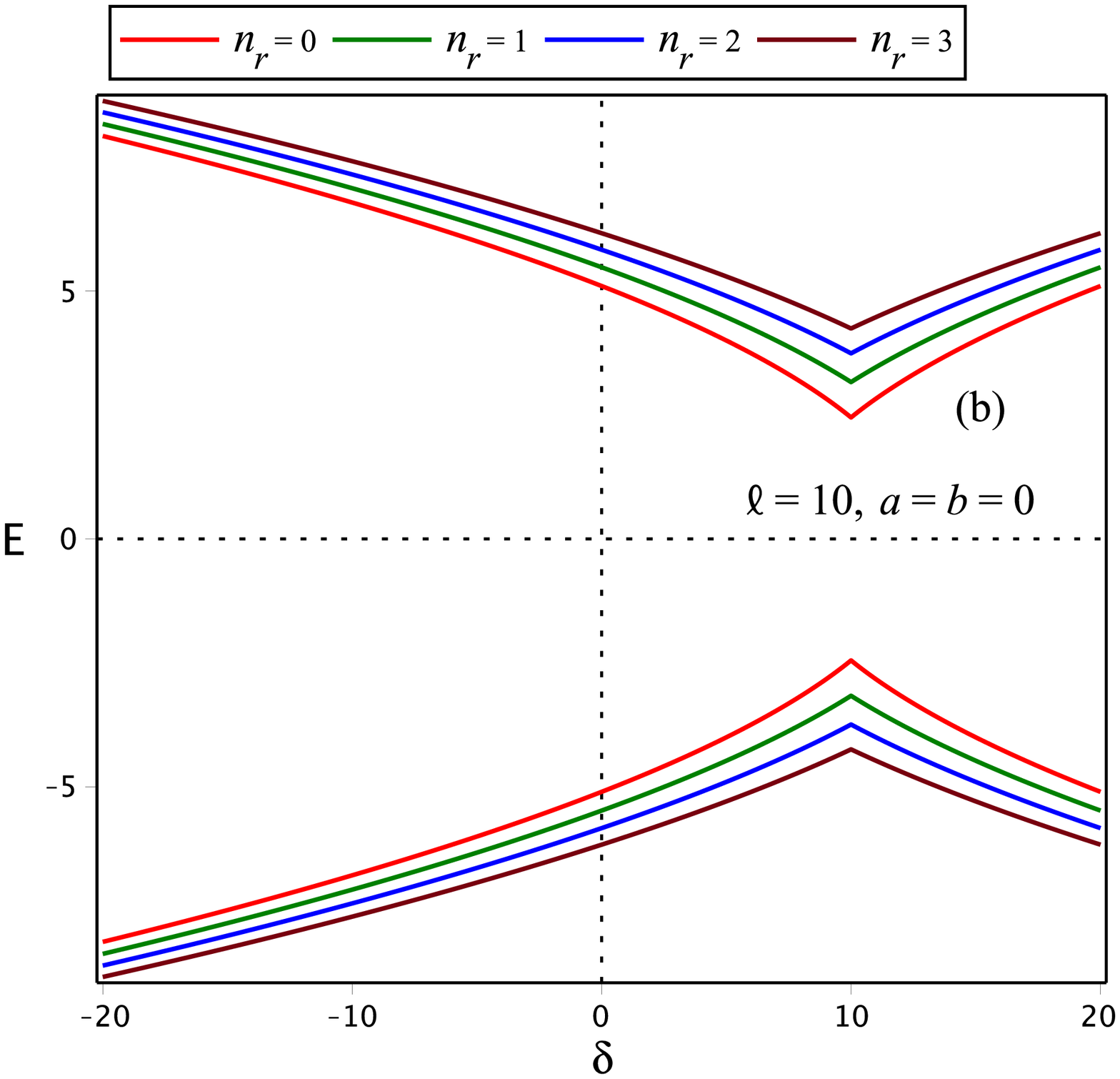}
\includegraphics[width=0.3\textwidth]{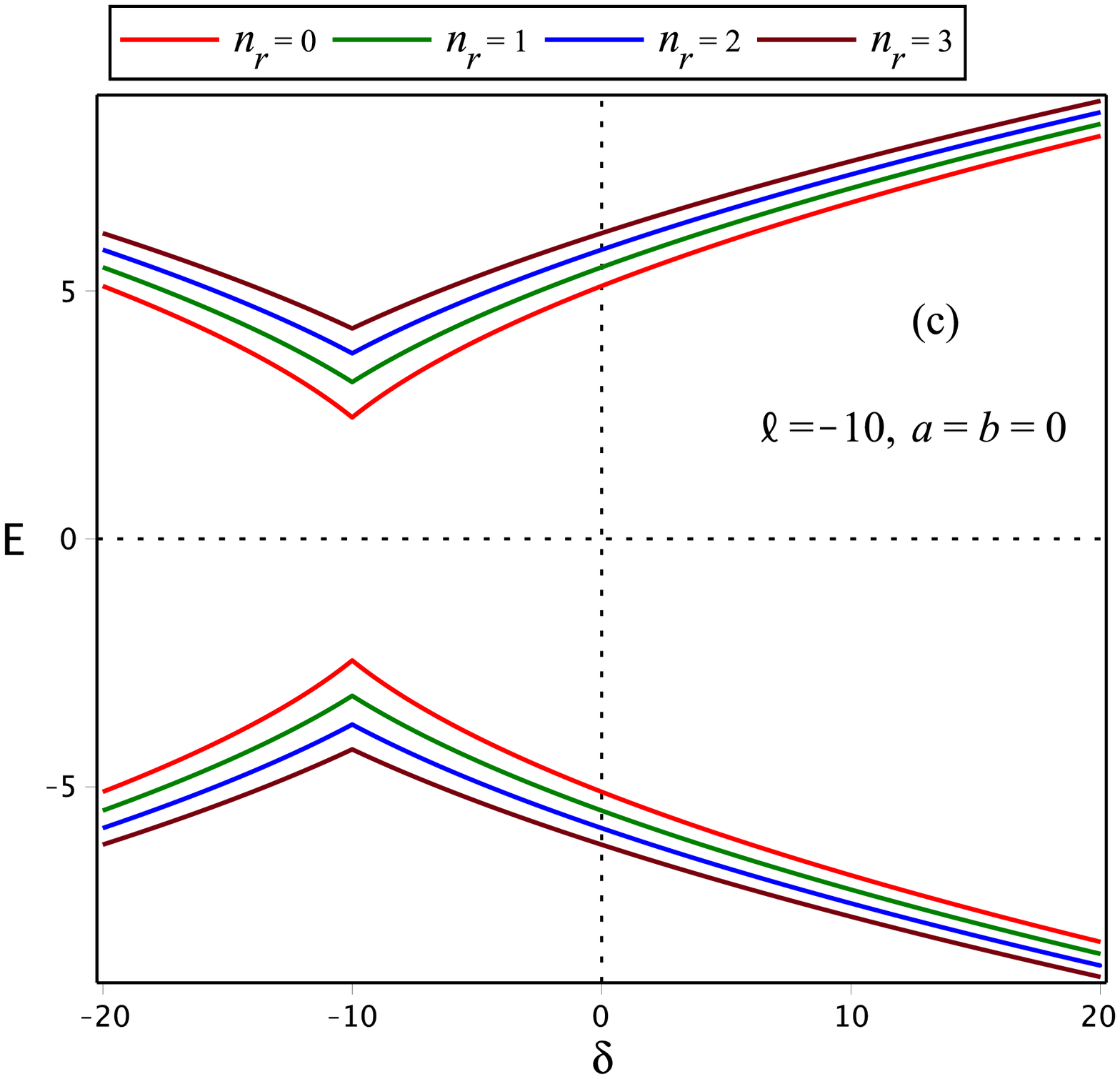}
\caption{\small 
{ We plot the energy levels of (\ref{energy eq 1}) versus the torsion parameter $\delta$, without the Cornell confinement, for $m=k_z=\eta=1$, $a=b=0$ and for (a) $\ell=0$, $n_r=0,1,2,3$, (b) $\ell=10$, $n_r=0,1,2,3$, and (c) $\ell=-10$, $n_r=0,1,2,3$.}}
\label{fig2}
\end{figure}%

The choice of $\gamma =2\left( 2n_{r}+1\right) +\alpha $ is not a random one but rather manifested by the fact that when $a=b=0$ the energies in (\ref{2D-lambda 1}) should naturally be recovered (this issue is emphasised in e.g., \cite{Mustafa1 2022,Ron 1995,Neto 2020}). However, in their comment on Vit\'{o}ria et al.'s \cite{R3} biconfluent Heun solution/polynomial, Neto et al. \cite{Neto 2020} have followed \cite{Ron 1995} and detailed the correct approach, which is very much in agreement with our treatment given in the Appendix below. This would also suggest that the result (33) of Vit\'{o}ria and Bakke \cite{R4} is valid for $n\geq 0$ and not for $n\geq 1$ (in this case they have lost all states with the quantum number $n_{r}=0$). The reader is advised to see also the detailed discussion on this quantum system and the use of $a_{n+1}=0$ condition that correlates $a$ with $b$, but at the same time introduces physically/mathematically unacceptable results, in the Appendix below.

At this point, one should be aware that this result (\ref{energy eq 1}), along with that in (\ref{S(r)-solution}), belong to the set of the so called \textit{conditionally} exactly solvable quantum mechanical problems. Moreover, it is obvious that for the case when $\tilde{\omega}=0=a$, the biconfluent Heun polynomial energies in (\ref{energy eq 1}) tragically fails to provide any information on the spectrum and/or the radial wave functions of a KG-Coulombic problem. Yet, instead of collapsing into the spectrum of
the KG-Coulombic problem, the reported spectrum (\ref{energy eq 1}) collapses into the free relativistic particle energies $E^{2}=m^{2}+k_{z}^{2}$. Nevertheless, we continue with such \textit{conditionally} exact solution (\ref{S(r)-solution}) and (\ref{energy eq 1}) and do our analysis. 

In Figures 1 and 2, we show the effect of dislocation related parameter $\delta $ on the energy levels of a confined KG-oscillator in Minkowski spacetime with space-like dislocation. We clearly observe that the first term under the square root of (\ref{energy eq 1}) determines the shifts in the energy levels at $\delta =\ell /k_{z}$, on the $\delta $-axis. That is, for negative $\ell $ values the shifts will be in the negative $\delta $ region, whereas for positive $\ell $ values the shifts will be in the
positive $\delta $ region. This would, in effect, manifestly yield energy levels crossings (i.e., occasional degeneracies, as shown in figures 1(a), 1(b), and 1(c), with the Cornell confinement). Moreover, in Figures 2(a), 2(b), and 2(c), we observe eminent energy levels clusterings when $\left\vert \delta \right\vert >>1$, for each value of the magnetic quantum number $\ell =0,1,2,\cdots $. These effects of the dislocation parameter on the energy levels of the confined KG-oscillator in Minkowski spacetime with
space-like dislocation are clear, therefore. 

\section{KG-particles in a pseudo-Minkowski spacetime with space-like dislocation admitting isospectrality and invariance with the KG-oscillators of (\ref{screw metric 1})}

Let metric (\ref{screw metric 1}) that describes Minkowski spacetime with space-like dislocation be transformed in such a way that%
\begin{equation}
ds^{2}\longrightarrow d\tilde{s}^{2}=-d\tilde{t}^{2}+d\tilde{r}^{2}+\tilde{r}%
^{2}d\tilde{\varphi}^{2}+\left( d\tilde{z}+\delta d\tilde{\varphi}\right)
^{2},  \label{screw metric 2}
\end{equation}%
where%
\begin{equation}
d\tilde{r}=\sqrt{g\left( r\right) }dr\Rightarrow\tilde{r}=\int \sqrt{g\left(
r\right) }dr=\sqrt{Q\left( r\right) }r,\,\,d\tilde{\varphi}=d\varphi ,\,\,d%
\tilde{z}=dz,\,\,d\tilde{t}=dt.  \label{PT-PDM}
\end{equation}%
This would in turn imply that%
\begin{equation}
\frac{d\tilde{r}}{dr}\Longrightarrow \sqrt{g(r)}=\sqrt{Q\left( r\right) }%
\left[ 1+\frac{Q^{\prime }\left( r\right) }{2Q\left( r\right) }r\right].
\label{Q-m relation}
\end{equation}%
This would govern the correlation between the positive-valued scalar multipliers $g\left( r\right) $ and $Q\left( r\right) $. In this case, our transformed metric (\ref{screw metric 2}) reads%
\begin{equation}
d\tilde{s}^{2}=-dt^{2}+g(r)\,dr^{2}+Q(r)\,r^2\,d\varphi^{2}+\left( dz+\delta
d\varphi\right) ^{2}.  \label{screw metric 2-1}
\end{equation}%
In this section, we shall show that all KG-particles in such pseudo-Minkowski spacetime with space-like dislocation (\ref{screw metric 2-1}) exactly inherit the quantum mechanical properties of the confined KG-oscillator in the Minkowski spacetime with space-like screw dislocation discussed in section 2 above. At this point, one should notice that our positive-valued scalar multiplier $g(r)$ should never converge to zero as $r\rightarrow \infty$, otherwise it would yield a catastrophic collapse of
the radial coordinate (any related coordinate in general) and consequently a catastrophic collapse of the quantum mechanical system at hand. In the classical mechanical language, $\tilde{r}=\sqrt{Q( r)}r \Rightarrow \dot{\tilde{r}}=d\tilde{r}/dt=\sqrt{g( r)}\dot{r}$ suggests that $Q(r)> 0$ and $g(r)>0$ so that the $0\leq (\tilde{r}, r)<\infty$. More details on origin of such transformation are given in (see e.g., \cite{Mustafa Phys.Scr. 2020} and related references cited therein). Then the covariant and contravariant metric tensors (with $f\left( r\right)=\tilde{r}^2 =Q\left( r\right) r^{2}$ for economy of notations) in this case, respectively, read%
\begin{equation}
\tilde{g}_{\mu \nu }=\left( 
\begin{tabular}{cccc}
$-1\smallskip $ & $\,0\,$ & $0$ & $\,0$ \\ 
$0$ & $g(r)\smallskip $ & $0$ & $0$ \\ 
$0$ & $\,0$ & $\,\left( f\left( r\right) +\delta ^{2}\right) \,$ & $\delta $
\\ 
$0$ & $0$ & $\delta $ & $1$%
\end{tabular}%
\right) \Longleftrightarrow \tilde{g}^{\mu \nu }=\left( 
\begin{tabular}{cccc}
$-1\smallskip $ & $\,0\,$ & $0$ & $0$ \\ 
$0$ & $\,\frac{1\smallskip }{g(r)}$ & $0$ & $0$ \\ 
$0$ & $\,0$ & $\,\frac{1}{f\left( r\right) }\,$ & $-\frac{\delta }{f\left(
r\right) }$ \\ 
$0$ & $0$ & $-\frac{\delta }{f\left( r\right) }\,$ & $\,-\left( 1+\frac{%
\delta ^{2}}{f\left( r\right) }\right) $%
\end{tabular}
\right) \text{ };\text{ \ }\det \left( \tilde{g}_{\mu \nu }\right) =-g\left(
r\right) f\left( r\right) .  \label{g-metric 2}
\end{equation}%
We now include the PDM KG-oscillator using the momentum operator (\ref{momentum op}) of Mirza et al.'s recipe \cite{Mirza 2004} and suggest that $\chi _{\mu }=\left( 0,\sqrt{g\left( r\right) f\left( r\right) },0\right)$ to accommodate a new set of KG-oscillators in the pseudo-spacetime with space-like dislocation settings. This would, in effect, transform the KG-oscillator equation (\ref{KG-oscillator 1}) into%
\begin{equation}
\left[ \frac{\partial _{r}}{\sqrt{g\left( r\right) f\left( r\right) }}\left( 
\sqrt{\frac{f\left( r\right) }{g\left( r\right) }}\partial _{r}\right)
-\partial _{t}^{2}+\frac{\partial _{\varphi }^{2}}{f\left( r\right) }+\left[
1+\frac{\delta ^{2}}{f\left( r\right) }\right] \partial _{z}^{2}-\frac{%
2\delta \partial _{z}\partial _{\varphi }}{f\left( r\right) }-2\eta -\eta
^{2}f\left( r\right) -(m+S\left( \tilde{r}\right) )^{2}\right] \Psi =0.
\label{KG-eq2}
\end{equation}%
Which upon the substitution%
\begin{equation}
\Psi \left( t,r,\varphi ,z\right) =\exp \left( i\left[ \ell \varphi
+k_{z}z-Et\right] \right) U\left( r\right) ,  \label{Psi 2}
\end{equation}%
and%
\begin{equation}
S(\tilde{r})=a \,\tilde{r}+\frac{b}{\tilde{r}}=a\,\sqrt{f(r)}+\frac{b}{\sqrt{%
f(r)}},  \label{ST(r)}
\end{equation}
yields%
\begin{equation}
\frac{\partial _{r}}{\sqrt{g\left( r\right) f\left( r\right) }}\left( \sqrt{%
\frac{f\left( r\right) }{g\left( r\right) }}\partial _{r}\right)
U\left(r\right) +\left[ \lambda -\frac{\tilde{\gamma}^{2}}{f\left( r\right) }%
-{\tilde{\omega}}^{2}f\left( r\right) -2m\,\left(a\sqrt{f(r)}+\frac{b}{\sqrt{%
f(r)}}\right)\right] U\left( r\right)=0 .  \label{U(r)-eq}
\end{equation}%
Where $\lambda$, $\tilde{\gamma}^2$, and ${\tilde{\omega}}^{2}$ are defined in (\ref{Cornell eq parameters}). Yet, the first term of (\ref{U(r)-eq}) can be rewritten, with $f\left( r\right)=\tilde{r}^{2} $ and $\partial _{\tilde{r}}=\frac{1}{\sqrt{g\left( r\right) }}\partial _{r}$, as%
\begin{equation}
\frac{1}{\sqrt{f\left( r\right) }}\frac{1}{\sqrt{g\left( r\right) }}\partial
_{r}\left( \sqrt{f\left( r\right) }\frac{1}{\sqrt{g\left( r\right) }}%
\partial _{r}\right) U\left( r\right) =\frac{1}{\tilde{r}}\frac{\partial }{%
\partial \tilde{r}}\left( \tilde{r}\,\frac{\partial }{\partial \tilde{r}}%
\right) U\left( \tilde{r}(r)\right) =\left( \frac{\partial ^{2}}{\partial 
\tilde{r}^{2}}+\frac{1}{\tilde{r}}\frac{\partial }{\partial \tilde{r}}%
\right) U\left( \tilde{r}(r)\right) .  \label{U(r)-R(r) 1}
\end{equation}%
To remove the first derivative we may define $U\left( \tilde{r}\right) = R\left( \tilde{r}\right) /\sqrt{\tilde{r}}$ to eventually imply%
\begin{equation}
\frac{d^{2}}{d\tilde{r}^{2}}R\left( \tilde{r}\right) +\left[ \lambda -\frac{%
\left( \tilde{\gamma}^{2}-1/4\right) }{\tilde{r}^{2}}-{\tilde{\omega}}^{2}%
\tilde{r}^{2}-2ma\tilde{r}-2m\frac{b}{\tilde{r}}\right] R\left( \tilde{r}%
\right) =0.  \label{R(tilde r) eq}
\end{equation}%
This equation is in the same form as that in (\ref{cornell-1 R(r) eq}) and they are, therefore, isospectral and invariant. Hence, (\ref{R(tilde r) eq}) inherits the energies reported in (\ref{energy eq 1}) and the radial eigenfunctions in (\ref{S(r)-solution}) but with $\tilde{r}$ replacing $r$. That is, in terms of the biconfluent Heun polynomials, the eigenfunctions would read%
\begin{equation}
\psi(\tilde{r})=\frac{R\left(\tilde{r}\right)}{\sqrt{\tilde{r}}} \sim 
\mathcal{\,}\tilde{r}^{\left\vert \tilde{\gamma}\right\vert }\,\exp \left( -%
\frac{\ \tilde{\omega}^{2}\tilde{r}^{2}+2am \tilde{r}}{2\,\tilde{\omega}}%
\right) \,H_{B}\left( 2\left\vert \tilde{\gamma}\right\vert ,\frac{2ma}{\ 
\tilde{\omega}^{3/2}},\frac{a^{2}m^{2}+\tilde{\lambda}\,\ \tilde{\omega}^{2}%
}{\ \tilde{\omega}^{3}},\frac{4mb}{\sqrt{\ \tilde{\omega}}},\sqrt{\ \tilde{%
\omega}}\tilde{r}\right) ,  \label{S(r)-solution1}
\end{equation}%
As long as $g(r)$ and $Q(r)$ (of $f(r)=\tilde{r}^2=Q(r)r^2$) are correlated through (\ref{Q-m relation}) and they are positive valued functions, then all KG-particles in the transformed pseudo-Minkowski spacetime with space-like dislocation metric (\ref{screw metric 2-1}) have identical energy spectra as the spectrum of the KG-oscillators in Minkowski spacetime with space-like dislocation metric (\ref{screw metric 1}), confined $S(r)\neq 0$ or unconfined $S(r)=0$, discussed in section 2 above.

\section{PDM KG-oscillators in Minkowski spacetime with space-like dislocation and confined/unconfined KG-oscillators}

We have mentioned that Mustafa and Algadhi \cite{Mustafa Algadhi 2019} have shown that an effective PDM-momentum operator is given by (\ref{PDM-op}). In this section, we shall use such PDM-momentum operator to describe \textit{metaphorically} PDM KG-particles in a spacetime with a space-like dislocation (\ref{screw metric 1}) and subject them to a Lorentz scalar potential $S(r)$. The corresponding inverse metric tensor $g^{\mu \nu }$ is readily given in (\ref{g-metric1}). Moreover, we shall use the assumption
that $m\left( \mathbf{r}\right) =m\left( r\right)=m\,g(r) $ (i.e., only radially dependent). Under such settings, the momentum operator in (\ref{momentum op}) would take the PDM form so that%
\begin{equation}
\tilde{p}_{\mu }\longrightarrow -i\partial _{\mu }+i\mathcal{F}_{\mu }\,%
\mathbf{\ ;}\,\,\text{ }\mathcal{F}_{\mu }=\left( 0,\mathcal{F}%
_{r},0,0\right) ,\,\, \mathcal{F}_{r}=\eta\, r+\frac{g^{\prime }\left(
r\right) }{4\,g\left( r\right) },  \label{PDM KG-oscillator momentum}
\end{equation}%
is used to construct the KG-oscillators with PDM in a spacetime with a space-like dislocation through%
\begin{equation}
\frac{1}{\sqrt{-g}}\left( \partial _{\mu }+\mathcal{F}_{\mu }\right) \left[ 
\sqrt{-g}g^{\mu \nu }\left( \partial _{\nu }-\mathcal{F}_{\nu }\right) \Psi %
\right] =\left( m+S\left( r\right) \right) ^{2}\Psi .  \label{PDM KG-eq}
\end{equation}%
This equation (\ref{PDM KG-eq}), with the contravariant metric tensors in (\ref{g-metric1}), would yield%
\begin{equation}
\left[ \frac{1}{r}\partial _{r}\,r\partial _{r}-\partial _{t}^{2}+\frac{1}{%
r^{2}}\partial _{\varphi }^{2}+\left( 1+\frac{\delta ^{2}}{r^{2}}\right)
\partial _{z}^{2}-\frac{2\delta }{r^{2}}\partial _{z}\partial _{\varphi }-%
\mathcal{F}_{r}^{^{\prime }}-\frac{\mathcal{F}_{r}}{r}-\mathcal{F}%
_{r}^{2}-(m+S\left( r\right) )^{2}\right] \Psi =0.  \label{PDM Psi-eq}
\end{equation}%
We may now use $\Psi \left( t,r,\varphi ,z\right) $ of (\ref{Psi 1}) to obtain%
\begin{equation}
R^{\prime \prime }\left( r\right) +\left[ \lambda -\frac{\left( \tilde{\ell}%
^{2}-1/4\right) }{r^{2}}-\eta^2\,r^2-2mS\left( r\right) -S\left( r\right)
^{2}+M\left( r\right) \right] R\left( r\right) =0  \label{PDM R(r)-eq}
\end{equation}%
where $\lambda =E^{2}-k_{z}^{2} -m^{2}-2\,\eta$, $\tilde{\ell}^{2}=\left(\ell -k_{z}\delta \right) ^{2}$, and%
\begin{equation}
\,M\left( r\right) =\frac{3}{16}\left( \frac{g^{\prime }\left( r\right) }{%
g\left( r\right) }\right) ^{2}-\frac{1}{4}\frac{g^{\prime \prime }\left(
r\right) }{g\left( r\right) }-\frac{g^{\prime }\left( r\right) }{%
4\,r\,g\left(r\right) }-\frac{g^{\prime }\left( r\right) }{%
2\,g\left(r\right) }\,\eta\, r.  \label{PDM M(r)}
\end{equation}

Obviously, for constant mass settings the dimensionless scalar multiplier is set equal 1, i.e., $g\left( r\right) =1$, and equation (\ref{PDM R(r)-eq}) collapses into that of (\ref{radial eq 1}) as should be. Yet, one should notice that when the KG-oscillator's effective frequency is off, i.e., $\eta=0$, equation (\ref{PDM R(r)-eq}) would describe KG-particles in Minkowski spacetime with a space-like dislocation, in general. To study the space-like dislocation effect on such PDM KG-particles, we choose three illustrative examples.

\subsection{Example 1: A PDM KG-oscillator from a dimensionless scalar multiplier $g(r)=exp(2\alpha r^2) \geq 0, \alpha\geq 0$ }

Let us start with a dimensionless scalar multiplier $g\left( r\right) =\,exp(2\alpha r^2)\geq0,\, \alpha\,\geq 0$ to imply that%
\begin{equation}
M\left( r\right) =-(\alpha^2+2\,\alpha\,\eta)\,r^{2}-2\alpha \,
\label{M(r) 1}
\end{equation}%
This would, with $\Omega=\alpha+\eta$, imply that equation (\ref{PDM R(r)-eq}) now reads%
\begin{equation}
R^{\prime \prime }\left( r\right) +\left[ \lambda_1 -\frac{\left( \tilde{\ell%
}^{2}-1/4\right) }{r^{2}}-\Omega ^{2}r^{2}-2mS\left( r\right) -S\left(
r\right) ^{2}\right] R\left( r\right) =0,  \label{radial eq 31}
\end{equation}%
where%
\begin{equation}
\lambda_1 =E^{2}-k_{z}^{2}-2\Omega -m^{2}\, ; \, \,\tilde{\ell}^{2}=\left(
\ell -k_{z}\delta \right) ^{2}.  \label{parametres 1}
\end{equation}%
Obviously, the effective angular frequency $\Omega=\alpha+\eta\geq 0$ of this model suggests that a KG-oscillator could also be a manifestation of a dimensionless scalar multiplier $g\left( r\right) =\, exp(2\alpha r^2),\, \alpha\,\geq 0$, when $\eta=0$. This is yet another way to come out with a KG-oscillator like model. Moreover, for a Cornell-type confining potential $S(r)$ (\ref{Cornell potential}) we obtain%
\begin{equation}
R^{\prime \prime }\left( r\right) +\left[ \tilde{\lambda}_1-\frac{\left( 
\tilde{\gamma}^{2}-1/4\right) }{r^{2}}-\tilde{\Omega}^{2}r^{2}-2mar-\frac{2mb%
}{r}\right] R\left( r\right) =0,  \label{cornell R(r) eq}
\end{equation}%
where%
\begin{equation}
\tilde{\lambda}_1=E^{2}-k_{z}^{2}-2\Omega -m^{2}-2ab\,;\,\, \tilde{\gamma}%
^{2}=\left( \ell -k_{z}\delta \right) ^{2}+b^{2}\,;\,\, \tilde{\Omega}%
^2=(\alpha+\eta)^2+a^2.  \label{Cornell eq parameters 3}
\end{equation}%
Now, $\tilde{\gamma}=\pm\sqrt{\left( \ell -k_{z}\delta \right) ^{2}+b^{2}}$ is the new irrational magnetic quantum number and $\tilde{\Omega}=\sqrt{\Omega^{2}+a^{2}}\geq0$ is our new effective oscillation frequency. This equation is in the same form as that in (\ref{cornell-1 R(r) eq}), and hence it admits similar forms of the eigenfunctions (\ref{S(r)-solution}) and energies (\ref{energy eq 1}) with $\tilde\Omega$ replaces $\tilde{\omega}$. That is,%
\begin{equation}
\psi(r)=\frac{R\left( r\right)}{\sqrt{r}} \sim \mathcal{\,}r^{\left\vert 
\tilde{\gamma}\right\vert }\,\exp \left( -\frac{\ \tilde{\Omega}%
^{2}r^{2}+2amr}{2\,\tilde{\Omega}}\right) \,H_{B}\left( 2\left\vert \tilde{%
\gamma}\right\vert ,\frac{2ma}{\ \tilde{\Omega}^{3/2}},\frac{a^{2}m^{2}+%
\tilde{\lambda}_1\,\ \tilde{\Omega}^{2}}{\ \tilde{\Omega}^{3}},\frac{4mb}{%
\sqrt{\ \tilde{\Omega}}},\sqrt{\ \tilde{\Omega}}r\right) ,
\label{S(r)-solution12}
\end{equation}%
and%
\begin{equation}
\frac{a^{2}m^{2}+\tilde{\lambda}_1\,\tilde{\Omega}^{2}}{\tilde{\Omega}^{3}}%
=2\left( 2n_{r}+\left\vert \tilde{\gamma}\right\vert +1\right)
\Longleftrightarrow \tilde{\lambda}_1=2\tilde{\Omega}\left(
2n_{r}+\left\vert \tilde{\gamma}\right\vert +1\right) -\frac{m^{2}a^{2}}{%
\tilde{\Omega}^{2}}  \label{lambda-S(r) 12}
\end{equation}%
In this case, we get the relation for the energy eigenvalues as%
\begin{equation}
E^{2}=2\left( \sqrt{(\alpha+\eta) ^{2}+a^{2}}\right) \left(
2n_{r}+\left\vert \sqrt{\left( \ell -k_{z}\delta \right) ^{2}+b^{2}}%
\right\vert +1\right) -\frac{m^{2}a^{2}}{(\alpha+\eta)^{2}+a^{2}}%
+2(\alpha+\eta) +k_{z}^{2}+m^{2}+2ab.  \label{energy eq 12}
\end{equation}%
Which for $\alpha=0$ retrieves the result in (\ref{S(r)-solution}) and (\ref{energy eq 1}).

\subsection{Example 2: A PDM KG-oscillator from a power law type dimensionless scalar multiplier $g(r)=A r^\sigma \geq 0$}

A power-law type dimensionless scalar multiplier $g\left( r\right)=Ar^{\sigma }\geq0$ would, through (\ref{PDM M(r)}), imply that%
\begin{equation}
M\left( r\right) =-\frac{\sigma ^{2}}{16r^{2}}-\frac{\sigma }{2}\eta .
\label{M(r) power-law}
\end{equation}%
\begin{figure}[!ht]  
\centering
\includegraphics[width=0.3\textwidth]{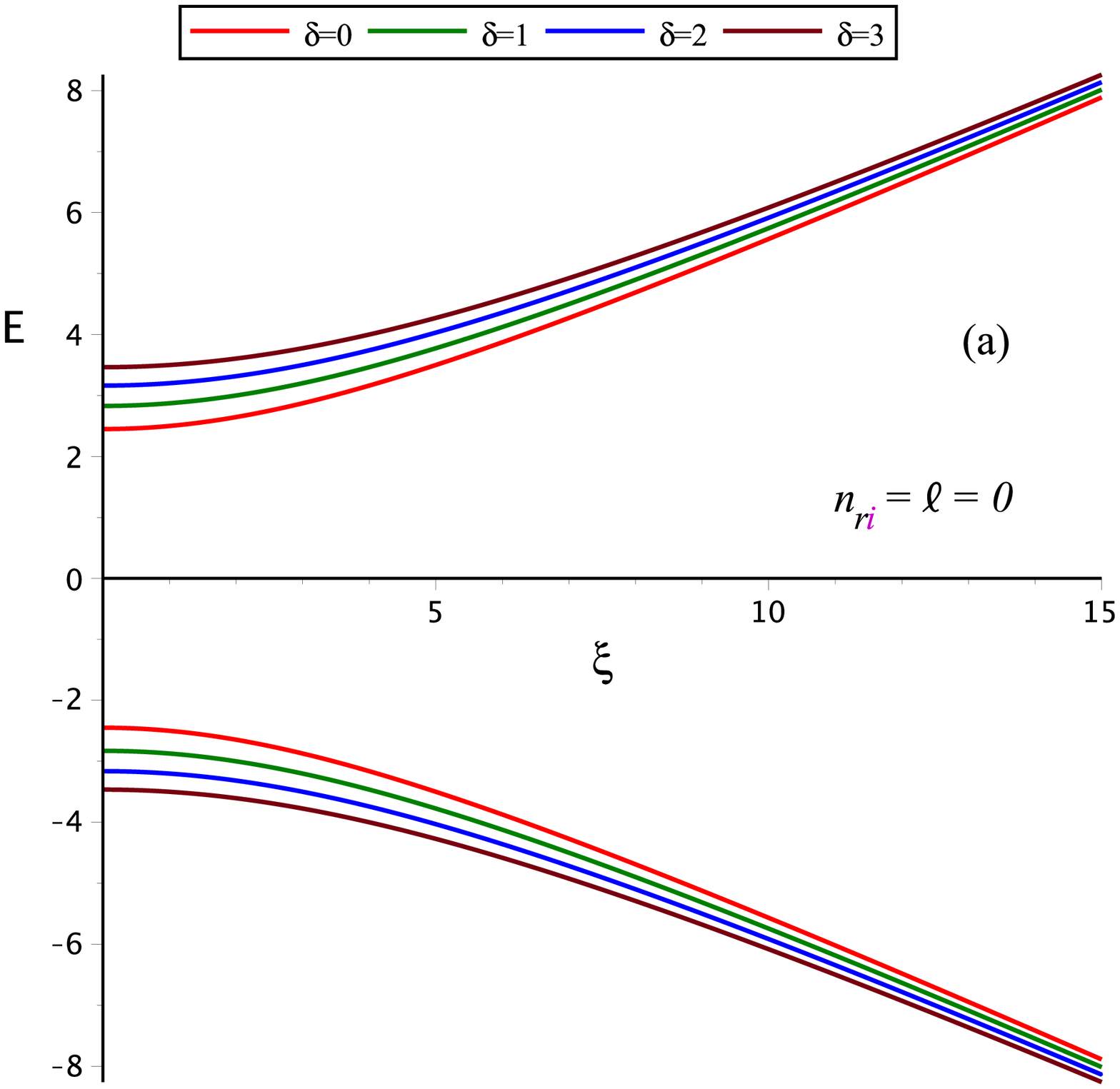}
\includegraphics[width=0.3\textwidth]{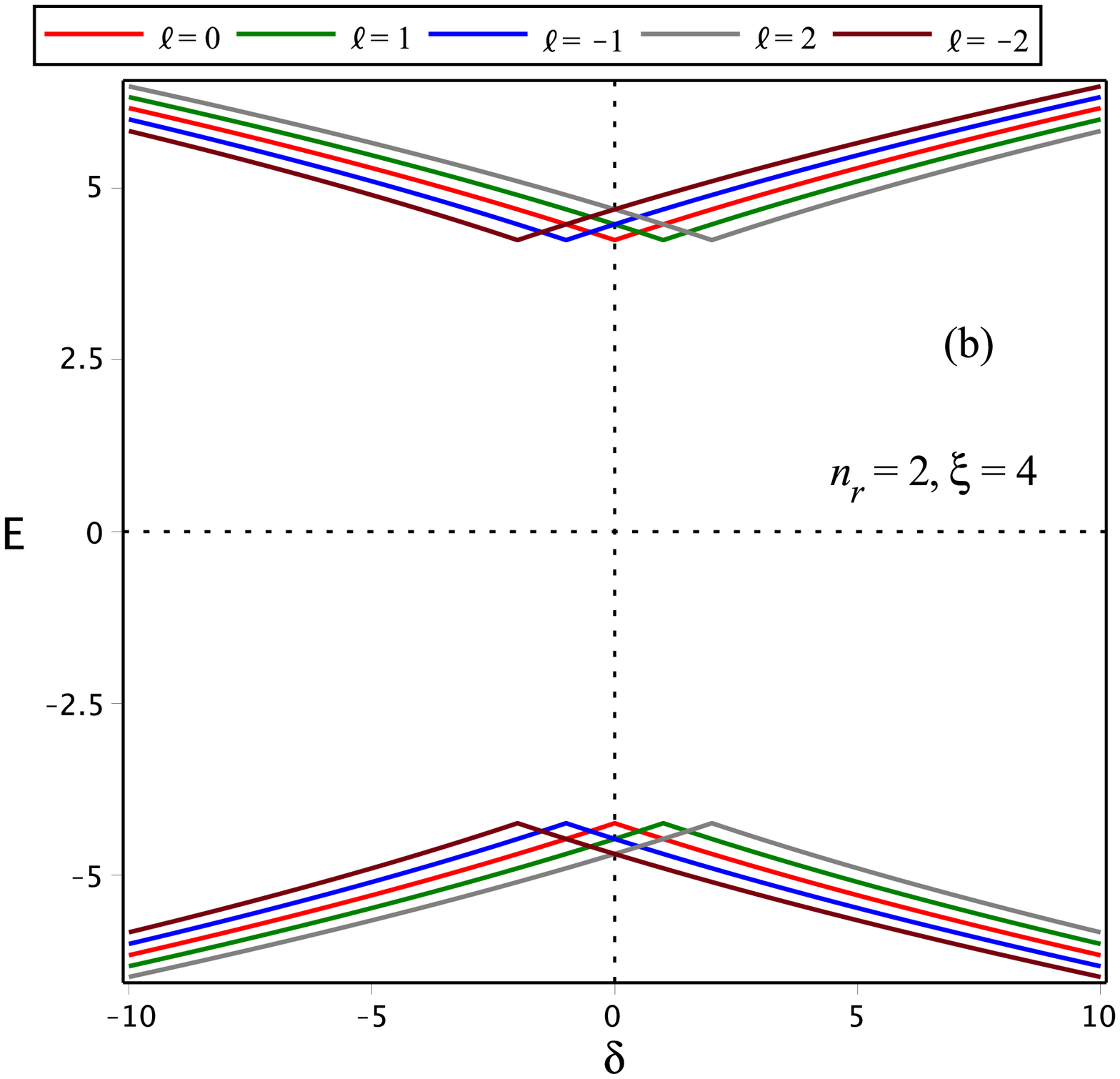}
\includegraphics[width=0.3\textwidth]{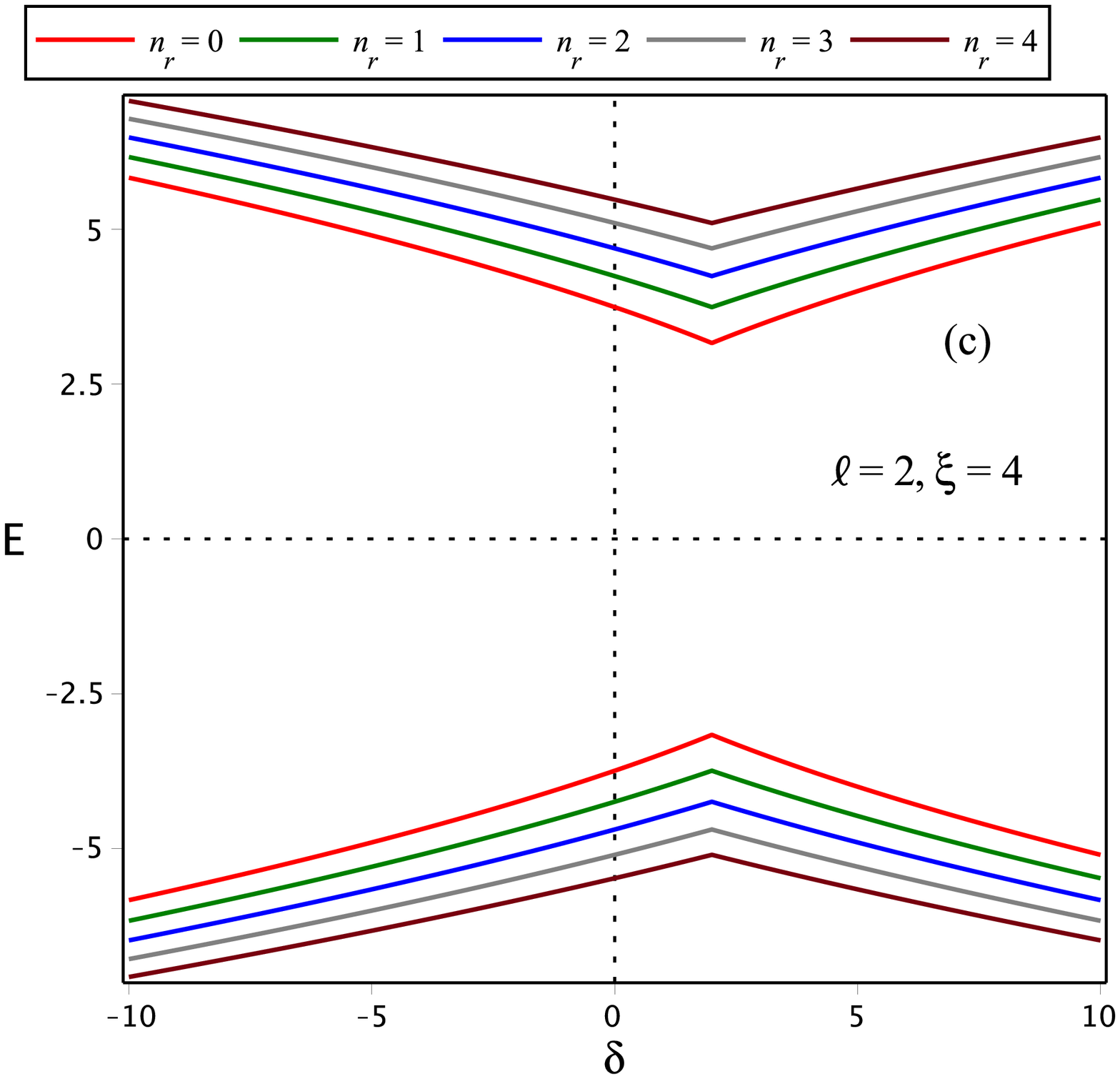}
\caption{\small 
{ We plot the energy levels (\ref{energy levels exponential}) of the exponentially growing PDM for $m=k_z=\eta=1$. We show in (a) the effect of the PDM parameter $\xi$ for $n_{r}=\ell=0$, (b) the effect of the torsion parameter $\delta$, for $n_r=2$, $\xi=4$, $\ell=0,\pm1,\pm2$, and (c) the effect of the torsion parameter $\delta$, for $\ell=2, \xi=4, n_r=0,1,2,3,4$.}}
\label{fig3}
\end{figure}%

Which, in turn, yields%
\begin{equation}
R^{\prime \prime }\left( r\right) +\left[ \mathcal{E}-\frac{\left( \zeta
^{2}-1/4\right) }{r^{2}}-\eta ^{2}r^{2}-2mS\left( r\right) -S\left( r\right)
^{2}\right] R\left( r\right) =0,  \label{R(r)-eq power-law}
\end{equation}%
where%
\begin{equation}
\mathcal{E}=E^{2}-k_{z}^{2}-2\eta -m^{2}-\frac{\sigma }{2}\eta \,\,;\,\zeta
^{2}=\left( \ell -k_{z}\delta \right) ^{2}+\frac{\sigma ^{2}}{16}.
\label{power-law parameters}
\end{equation}%
With $S\left( r\right) =0$, this equation resembles that of the two-dimensional radial Schr\"{o}dinger oscillator discussed in section 2 (namely, equations (\ref{radial eq 1}), (\ref{2D-lambda 1}), and (\ref{radial wave 1})) and admits eigenvalues%
\begin{equation}
\mathcal{E}=2\eta \left( 2n_{r}+\left\vert \zeta \right\vert +1\right)
\Longleftrightarrow E^{2}=2\eta \left( 2n_{r}+\left\vert \sqrt{\left( \ell
-k_{z}\delta \right) ^{2}+\frac{\sigma ^{2}}{16}}\right\vert +2\right)
+k_{z}^{2}+m^{2}+\frac{\sigma }{2}\eta ,  \label{power-law E}
\end{equation}%
and radial eigenfunctions%
\begin{equation}
R\left( r\right) \sim r^{\left\vert \zeta \right\vert +1/2}\exp \left( -%
\frac{\eta r^{2}}{2}\right) L_{n_{r}}^{\left\vert \zeta \right\vert }\left(
\eta r^{2}\right) \Longleftrightarrow \psi \left( r\right) \sim
r^{\left\vert \zeta \right\vert }\exp \left( -\frac{\eta r^{2}}{2}\right)
L_{n_{r}}^{\left\vert \zeta \right\vert }\left( \eta r^{2}\right) .
\label{R(r) power-law}
\end{equation}

Let us now consider the PDM KG-oscillators confined in the Cornell-type potential of (\ref{Cornell potential}). This would, in effect, imply that equation (\ref{R(r)-eq power-law}) be rewritten as%
\begin{equation}
R^{\prime \prime }\left( r\right) +\left[ \mathcal{\tilde{E}}-\frac{\left( 
\tilde{\zeta}^{2}-1/4\right) }{r^{2}}-\tilde{\omega}^{2}r^{2}-2mar-\frac{2mb%
}{r}\right] R\left( r\right) =0,  \label{power law confined eq}
\end{equation}%
where%
\begin{equation}
\mathcal{\tilde{E}=E}-2ab\,;\,\,\text{ }\tilde{\zeta}^{2}=\zeta
^{2}+b^{2}\,;\,\,\tilde{\omega}^{2}=\eta ^{2}+a^{2}.
\label{power law parameters}
\end{equation}%
Again, this equation is in the form of (\ref{cornell-1 R(r) eq}), and hence it admits similar forms of the eigenfunctions (\ref{S(r)-solution}) and energies (\ref{energy eq 1}), with $\tilde{\zeta}$ replacing $\tilde{\ell}$, respectively, %
\begin{equation}
\psi \left( r\right) \sim \mathcal{\,}r^{\left\vert \tilde{\zeta}\right\vert
}\,\exp \left( -\frac{\ \tilde{\omega}^{2}r^{2}+2amr}{2\,\tilde{\omega}}%
\right) \,H_{B}\left( 2\left\vert \tilde{\zeta}\right\vert ,\frac{2ma}{\ 
\tilde{\omega}^{3/2}},\frac{a^{2}m^{2}+\mathcal{\tilde{E}}\,\ \tilde{\omega}%
^{2}}{\ \tilde{\omega}^{3}},\frac{4mb}{\sqrt{\ \tilde{\omega}}},\sqrt{\ 
\tilde{\omega}}r\right) ,  \label{HeunB power law}
\end{equation}%
and%
\begin{equation}
\frac{a^{2}m^{2}+\mathcal{\tilde{E}}\,\tilde{\omega}^{2}}{\tilde{\omega}^{3}}%
=2\left( 2n_{r}+\left\vert \tilde{\zeta}\right\vert +1\right)
\Longleftrightarrow \mathcal{\tilde{E}}=2\tilde{\omega}\left(
2n_{r}+\left\vert \tilde{\zeta}\right\vert +1\right) -\frac{m^{2}a^{2}}{%
\tilde{\omega}^{2}}  \label{HeunB parameters power law confined}
\end{equation}%
In this case, we get the relation for the energy eigenvalues as%
\begin{equation}
E^{2}=2\left( \sqrt{\eta ^{2}+a^{2}}\right) \left( 2n_{r}+\left\vert \sqrt{%
\left( \ell -k_{z}\delta \right) ^{2}+\frac{\sigma ^{2}}{16}+b^{2}}%
\right\vert +1\right) -\frac{m^{2}a^{2}}{\eta ^{2}+a^{2}}+2\eta
+k_{z}^{2}+m^{2}+2ab+\frac{\sigma }{2}\eta .
\label{nergies power law confined}
\end{equation}%
Obviously, such energy levels inherit the behavior of those of (\ref{energy eq 1}) discussed in section 2. That is, one may rewrite this energy equation
as%
\begin{equation}
E^{2}=2\tilde{\eta}\left( 2n_{r}+\left\vert \sqrt{\left( \ell -k_{z}\delta
\right) ^{2}+\tilde{b}^{2}}\right\vert +1\right) -\frac{m^{2}a^{2}}{\tilde{%
\eta}^{2}}+\left( 2+\frac{\sigma }{2}\right) \eta +k_{z}^{2}+m^{2}+2ab.
\label{E^2 inherited}
\end{equation}%
where $\tilde{\eta}=\sqrt{\eta ^{2}+a^{2}}$ and $\tilde{b}^{2}=b^{2}+\sigma^{2}/16$, to observe that similar trends of behavior.

\subsection{Example 3: A PDM KG-oscillator with a dimensionless scalar multiplier $g(r)=exp(\xi r) \geq 0$}

An exponentially growing dimensionless scalar multiplier $g\left( r\right)=\exp \left( \xi r\right) \geq 0$ would yield%
\begin{equation}
M\left( r\right) =-\frac{\xi ^{2}}{16}-\frac{1}{4}\frac{\xi }{r}-\frac{1}{2}%
\xi \eta r.  \label{M(r) PDM exponential}
\end{equation}%
Consequently, the PDM KG-oscillator's equation (\ref{PDM R(r)-eq}), with $S(r)=0$, reads%
\begin{equation}
R^{\prime \prime }\left( r\right) +\left[ \Sigma -\frac{\left( \tilde{\ell}%
^{2}-1/4\right) }{r^{2}}-\eta ^{2}r^{2}-\frac{1}{4}\frac{\xi }{r}-\frac{1}{2}%
\xi \eta r-2mS\left( r\right) -S\left( r\right) ^{2}\right] R\left( r\right)
=0,  \label{R(r) PDM-eq exponential}
\end{equation}%
where $\Sigma =E^{2}-k_{z}^{2}-2\eta -m^{2}-\xi ^{2}/16$, and $\tilde{\ell}^{2}=\left( \ell -k_{z}\delta \right) ^{2}$. It is clear that a Cornell-type confinement (i.e., $\xi /4r+\xi \eta r/2$) is introduced as a byproduct of the dimensionless scalar multiplier at hand. We, therefore, continue with $S(r)=0.$ This equation (\ref{R(r) PDM-eq exponential}) , with $S(r)=0$ and following the same procedure as that in the above examples, admits a solution in the form of biconfluent Heun polynomials%
\begin{equation}
\psi \left( r\right) \sim \mathcal{\,}r^{\left\vert \tilde{\ell}\right\vert
}\,\exp \left( -\frac{1}{2}\eta r^{2}-\frac{1}{4}\xi r\right) \,H_{B}\left(
2\left\vert \tilde{\ell}\right\vert ,\frac{\xi }{2\sqrt{\eta }},\frac{\xi
^{2}+16\Sigma }{16\eta },\frac{\xi }{2\sqrt{\ \eta }},\sqrt{\ \eta }r\right)
.  \label{R(r) exponential}
\end{equation}%
Hence, the corresponding energy levels are given by%
\begin{equation}
\frac{\xi ^{2}+16\Sigma }{16\eta }=2\left( 2n_{r}+\left\vert \tilde{\ell}%
\right\vert +1\right) \Longrightarrow E^{2}=2\,\eta \left( 2n_{r}+\left\vert
\ell -k_{z}\delta \right\vert +2\right) +k_{z}^{2}+m^{2}+\frac{1}{4}\eta \xi
^{2}.  \label{energy levels exponential}
\end{equation}%
The energy levels are shown in Figure 3. In Figure 3(a), we show the energy levels against the PDM parameter $\xi >0$ and observe eminent clustering of the energy levels as $\xi $ grows up, but no energy levels crossing are found feasible. On the other hand, the space-like dislocation parameter's effect on the energy levels, for some fixed values of $\xi $, maintains the same trend of behavior as that associated with (\ref{energy eq 1}) and
discussed in section 2.

\section{Concluding remarks}

In this work, we have studied the KG-oscillator in Minkowski spacetime with a space-like dislocation. We have started with KG-oscillators confined in a Cornell-type Lorentz scalar potential and discussed the dislocation effect on their conditionally exact energy levels. We observed that the space-like dislocation shifts the energy levels along the dislocation parameter $\delta$-axis by $\delta =\ell /k_{z};\,\ell =0,\pm 1,\pm 2,\cdots $ (documented in Figures 1(b), 1(c), 2(b), 2(c), 3(b), and 3(c)). That is, for $\ell =-|\ell|$ values, the shifts are in the direction of negative $\delta $- region, whereas for $\ell =+|\ell|$ values the shifts are in the direction of positive $\delta $ region. This in turn manifestly resulted in energy levels crossings (as shown in figures 1(b), 1(c), and 3(b)). Moreover, in Figures 2(a), 2(b), and 2(c), we have observed eminent energy levels clusterings when $\left\vert \delta \right\vert >>1$, for each value of the magnetic quantum number $\ell =0,\pm 1,\pm 2,\cdots $. We have reported, in section 3, a set of KG-particles in a pseudo-Minkowski spacetime with space-like dislocation admitting isospectrality and invariance with the confined KG-oscillators in Minkowski spacetime with a space-like dislocation. Such KG-particles are found to inherit the same effects discussed above.

We have used, in section 4, the argument that the momentum operator for PDM-particles (metaphorically speaking) is given by (\ref{PDM-op}) \cite{Mustafa Algadhi 2019} and yields a von Roos \cite{von Roos} kinetic energy operator (\ref{PDM-kinetic op}) with the so called MM-ordering (e.g., Mustafa and Mazharimousavi's ordering \cite{Mustafa 2020,Mustafa Habib 2007}). Such PDM-particles are studied in the context of KG-equation in Minkowski spacetime with a space-like dislocation background. Hence, the metaphoric
notion \textit{PDM KG-particles} is adopted in the process. The effect of space-like dislocation on energy levels of such KG-particles is reported through three illustrative examples, (i) a PDM KG-oscillator from a dimensionless scalar multiplier $g\left( r\right) =\, exp(2\alpha r^2)\geq0,\, \alpha\,\geq 0$, (ii) a PDM KG-oscillator from a power law type dimensionless scalar multiplier $g\left( r\right) =Ar^{\sigma }\geq0$, and (iii) a PDM KG-oscillator in a Cornnell-type confinement with a dimensionless scalar multiplier $g\left( r\right) =\exp \left( \xi r\right)\geq0$ . For the PDM KG-oscillators of (i) and (ii), the energy levels are shown to have similar trends of behavior as those of (\ref{energy eq 1}) discussed in section 2. Whereas, for the PDM KG-oscillator in (iii), we found that such PDM setting introduces a Cornell-like confinement as its
own byproduct. Hereby, obvious clustering of the energy levels are observed, as the PDM parameter $\xi $ grows up, but no energy levels crossing are found feasible for a fixed space-like dislocation parameter $\delta $ value (documented in figure 3(a)). Moreover, the effect of the space-like dislocation parameter $\delta $ on the energy levels, for a fixed PDM parameter $\xi $, is found to maintain the same trend of behavior as that associated with (\ref{energy eq 1}) and discussed in section 2.

Finally, the current methodical proposal may very well be extended to cover a more general case of PDM KG-particles and PDM Dirac-particles in different spacetime backgrounds with topological defects. In our opinion, the metaphoric PDM concept for relativistic particles should follow the procedure described in the current methodical proposal, and not through the assumption that $m\longrightarrow m+S\left( r\right)=m(r) $ (as in, e.g., \cite{Mustafa Habib 2008,Mustafa Habib1 2007,Vitoria Bakke 2016}). To the best of our knowledge, such a PDM KG-oscillator in Minkowski spacetime with space-like dislocation methodical proposal has never been reported elsewhere.

\section{Appendix: On the solution of the Schr\"{o}dinger oscillator in a
Cornell-type potential}

Let us rewrite (\ref{cornell-1 R(r) eq}) in terms of $\psi (r)$ so that%
\begin{equation}
\psi ^{\prime \prime }\left( r\right) +\frac{1}{r}\psi ^{\prime }\left(
r\right) +\left[ \tilde{\lambda}-\frac{\tilde{\gamma}^{2}}{r^{2}}-\tilde{%
\omega}^{2}r^{2}-A\,r-\frac{B}{r}\right] \psi \left( r\right)
=0;\,\,A=2ma,\,\,B=2mb,\ \tilde{\omega}^{2}=\eta ^{2}+a^{2}  \label{A1}
\end{equation}%
and define%
\begin{equation}
\psi (r)=r^{|\tilde{\gamma}|}\,\exp \left[ -\frac{\tilde{\omega}\,r^{2}}{2}-%
\frac{A\,r}{2\,\tilde{\omega}}\right] \,\,H(r),  \label{A2}
\end{equation}%
to imply%
\begin{equation}
r\,H^{\prime \prime }\left( r\right) +\left[ 1+2|\tilde{\gamma}|-2\,\tilde{%
\omega}\,r^{2}-\frac{A}{\tilde{\omega}}r\right] H^{\prime }(r)+\left(
\lambda ^{\prime }+\omega ^{\prime }r\right) \,H(r)=0.  \label{A3}
\end{equation}%
Where%
\begin{equation}
\lambda ^{\prime }=-\frac{A}{2\,\tilde{\omega}}\gamma ^{\prime
}-B,\,\,\gamma ^{\prime }=1+2\left\vert \tilde{\gamma}\right\vert
,\,\,\omega ^{\prime }=\frac{A^{2}}{4\,\tilde{\omega}^{2}}+\tilde{\lambda}-\,%
\tilde{\omega}\gamma ^{\prime }-\,\tilde{\omega}.  \label{A4}
\end{equation}%
With%
\begin{equation}
H(r)=\sum\limits_{j=0}^{\infty }a_{j}\,r^{j},  \label{A5}
\end{equation}%
in (\ref{A3}) would result%
\begin{equation}
\sum\limits_{j=0}^{\infty }\,\left\{ a_{j+2}\,\left[ (j+2)(j+\gamma ^{\prime
}+1)\right] +a_{j+1}\,\left[ \lambda ^{\prime }-\frac{A}{\tilde{\omega}}%
\,(j+1)\right] +a_{j}[\omega ^{\prime }-2\,\tilde{\omega}\,j]\,\right\}
\,r^{j+1}+\gamma ^{\prime }\,a_{1}+\lambda ^{\prime }\,a_{0}=0.  \label{A6}
\end{equation}%
Which in effect implies that%
\begin{equation}
\gamma ^{\prime }\,a_{1}+\lambda ^{\prime }\,a_{0}=0\Rightarrow a_{1}=-\frac{%
\lambda ^{\prime }}{\gamma ^{\prime }}\,;\,\,a_{0}=1.  \label{A7}
\end{equation}%
and%
\begin{equation}
a_{j+2}\,(j+2)(j+\gamma ^{\prime }+1)-a_{j+1}\,\left[ \frac{A}{\tilde{\omega}%
}\,(j+1)-\lambda ^{\prime }\right] +a_{j}\,[\omega ^{\prime }-2\,\tilde{%
\omega}\,j])=0\,;\,\,j=-1,0,1,2,\cdots ,\,a_{-1}=0.  \label{A8}
\end{equation}%
In order for the biconfluent Heun series to become a polynomial of degree $n\geq 0$, we truncate the power series by requiring that for $j=n$, $a_{n+1}=0$ and $a_{n+2}=0$. Consequently (\ref{A8}) would allow one to write%
\begin{equation}
a_{n}\,(\omega ^{\prime }-2\,\tilde{\omega}\,n)=0\,\Rightarrow \omega
^{\prime }=2\,\tilde{\omega}\,n\Rightarrow \tilde{\lambda}=2\,\tilde{\omega}%
\,(n+|\tilde{\gamma}|+1)-\frac{A^{2}}{4\,\tilde{\omega}^{2}};\,\,n\geq 0.
\label{A9}
\end{equation}%
In this case, following (\ref{A8}) we retrieve (\ref{A7}) for $j=-1$ and get, respectively, for $j=0,1,\cdots $%
\begin{equation}
a_{2}=\frac{1}{2\,(1+\gamma ^{\prime })}\left[ (\frac{A}{\tilde{\omega}}%
-\lambda ^{\prime })\,a_{1}-\omega ^{\prime }\right]   \label{A10}
\end{equation}%
\begin{equation}
a_{3}=\frac{1}{3\,(2\,|\tilde{\gamma}|+3)}\left[ (\frac{2\,A}{\tilde{\omega}}%
+\lambda ^{\prime })\,a_{2}-(\omega ^{\prime }-2\tilde{\omega})\,a_{1}\right]
\label{A11}
\end{equation}%
and%
\begin{equation}
a_{j+1}=\tilde{A}_{j-1\,}a_{j}+\tilde{B}_{j-1}\,a_{j-1};\,j=0,1,\cdots
,n,\,a_{-1}=0,  \label{A11-1}
\end{equation}%
where%
\begin{equation}
\;\tilde{A}_{j}=\frac{\frac{A}{\tilde{\omega}}\left( j+\frac{\gamma ^{\prime
}}{2}+1\right) +B}{(j+2)(j+\gamma ^{\prime }+1)},\;\tilde{B}_{j}=\frac{2%
\tilde{\omega}\left( j-n\right) }{(j+2)(j+\gamma ^{\prime }+1)}.
\label{A11-2}
\end{equation}%
Hereby, the recursion relation (\ref{A11-1}) along with (\ref{A11-2}) would identify the relations between $a_{j}^{\prime }s$ for $0\leq j\leq n$. Moreover, we demand that $\,a_{n+1}=0$ for $\forall j>n$. This would, using (\ref{A11-1}) and (\ref{A11-2}), allow us to obtain%
\begin{equation}
a_{n+1}=0=\tilde{A}_{n-1\,}a_{n}+\tilde{B}_{n-1}\,a_{n-1}\Longrightarrow
a_{n}\left[ \frac{A}{2\tilde{\omega}}\left( 2n+\gamma ^{\prime }\right) +B%
\right] =2\tilde{\omega}\,a_{n-1}.  \label{A11-3}
\end{equation}%
It is obvious that this equation would correlate $A$ and $B$ for each value of \ the truncation order $n$. For example, for%
\begin{eqnarray}
n &=&0\Longrightarrow A=-\frac{2\tilde{\omega}B}{\gamma ^{\prime }}%
,  \label{A11-4} \\
n &=&1\Longrightarrow \frac{A^{2}}{2\tilde{\omega}}\gamma ^{\prime }\left(
2+\gamma ^{\prime }\right) +2AB\left( 1+\gamma ^{\prime }\right) +2\tilde{%
\omega}B^2=\, 4\tilde{\omega}^{2}\gamma ^{\prime },  \label{11-5}
\end{eqnarray}%
and so on. This means that for every $n$ value we have a different correlation between $A=A(n,B)$ and $B=B(n,A)$ (which are in exact accord with those reported in (9) of Fern\'{a}ndez \cite{R6}). Under such sever restrictions on the parameters of the Cornell-type potential, one would recast the energy levels of (\ref{A9}) as%
\begin{equation}
\tilde{\lambda}=2\,\sqrt{\eta ^{2}+\frac{A^{2}}{4m}}\,(n+|\tilde{\gamma}|+1)-%
\frac{A^{2}}{4\,\left( \eta ^{2}+\frac{A^{2}}{4m}\right) };\,\,n\geq
0,\,A=A(n,B).  \label{A11-6}
\end{equation}

However, in order to retrieve the results of (\ref{2D-lambda 1}) we set $A=0=B\Longrightarrow \forall A(n,B)^{\prime }s=0$ in (\ref{A11-6}) and compare the two equations to come out with the correlation between the truncation order $n$ and the radial quantum number $n_{r}$ so that $n=2n_{r}\geq 0$. Therefore,%
\begin{equation}
\omega _{n_{r}}^{\prime }=4\,\tilde{\omega}\,n_{r}\Rightarrow \tilde{\lambda}%
=2\,\sqrt{\eta ^{2}+\frac{A^{2}}{4m}}\,(2\,n_{r}+|\tilde{\gamma}|+1)-\frac{%
A^{2}}{4\,\left( \eta ^{2}+\frac{A^{2}}{4m}\right) },  \label{A13}
\end{equation}%
to represent the eigenvalues of (\ref{cornell-1 R(r) eq}). At this point, one should notice that the condition of truncation of the biconfluent Heun series into a biconfluent Heun polynomial of degree $n=2n_{r}\geq 0$ is not violated. This is not a new practice. It has been discussed in \cite{Mustafa2 2022,Mustafa1 2022,Ron 1995,Neto 2020,R2,R5,R6,R7}. Obviously, moreover, this result shows that the biconfluent Heun polynomial solution discussed above is one of the so called conditionally exact solutions and not \textit{"the exact solution"} for (\ref{A1}). In this case, we rewrite our biconfluent Heun polynomial of degree $n=2\,n_{r}$ as%
\begin{equation}
H(r)=\sum\limits_{j=0}^{n=2\,n_{r}}\,a_{j}\,r^{j}\Rightarrow
H_{n_{r}}(r)=1+\sum\limits_{j=1}^{n=2\,n_{r}}\,a_{j}\,r^{j}.  \label{A14}
\end{equation}%
This polynomial is, in fact, responsible for the nodes in the corresponding radial wave function $\psi (r)$.

In this appendix section, we have followed, more or less, the usual procedure followed by many authors (e.g., \cite{Ron 1995,Neto 2020,R1,R2,R3,R4,R5,R6,R7} and references cited therein). In fact, we have very closely followed Fern\'{a}ndez \cite{R6,R7} to work out the above results. Fern\'{a}ndez \cite{R6,R7} has very carefully and righteously
detailed the most misunderstood conditionally-exact solvability of the model above, related to the three terms recursion relation (\ref{A11-1}). The result reported in (\ref{A13}) have derived some authors to conclude/claim that there exist some quantization recipe for the parameter $A=A(n,B)$ (hence, $B=B(n,B)$) mandated by the condition 
$a_{n+1}=a_{2n_{r}+1}=0$ in (\ref{A11-3}). However, if we put this above procedure to the test, we may then pin point the problem associated with such assumptions.

Let us consider that $B=0$ and $\left\vert \tilde{\gamma}\right\vert =\left\vert \ell -k_{z}\delta \right\vert =1/2$ in (\ref{A1}) and consequently (\ref{A1}), with $\psi \left( r\right) =R\left( r\right) /\sqrt{r}$, now reads%
\begin{equation}
R^{\prime \prime }\left( r\right) +\left[ \tilde{\lambda}-\tilde{\omega}%
^{2}r^{2}-A\,r\right] R\left( r\right) =0;\,\,A=2ma,\,\,B=2mb,\ \tilde{\omega%
}^{2}=\eta ^{2}+a^{2},\tilde{\lambda}=E^{2}-k_{z}^{2}-2\eta -m^{2}
\label{A15}
\end{equation}%
where the effect of the central repulsive/attractive core $\left( \tilde{\gamma}^{2}-1/4\right) /r^{2}$ is removed. Clearly, this equation resembles a shifted-harmonic oscillator and reduces to%
\begin{equation}
R^{\prime \prime }\left( r\right) +\left[ \Lambda -\tilde{\omega}^{2}\left(
r+\zeta \right) ^{2}\right] R\left( r\right) =0;\,\Lambda =\tilde{\lambda}\,+%
\frac{A^{2}}{4\tilde{\omega}^{2}},\,\zeta =\frac{A}{2\tilde{\omega}^{2}},
\label{A16}
\end{equation}%
that can be rewritten with $\tilde{r}=$ $r+\zeta $ as%
\begin{equation}
R^{\prime \prime }\left( \tilde{r}\right) +\left[ \Lambda -\tilde{\omega}^{2}%
\tilde{r}^{2}\right] R\left( \tilde{r}\right) =0.  \label{A17}
\end{equation}%
This equation denotes a radial harmonic Schr\"{o}dinger oscillator, without the central repulsive/attractive core $\left( \tilde{\gamma}^{2}-1/4\right)/r^{2}$, that admits the exact textbook eigenvalues%
\begin{equation}
\Lambda =2\tilde{\omega}\left( 2n_{r}+\frac{3}{2}\right) \Longrightarrow 
\tilde{\lambda}\,=2\tilde{\omega}\left( 2n_{r}+\frac{3}{2}\right) -\frac{%
A^{2}}{4\tilde{\omega}^{2}}\Longrightarrow E=\pm \left[ k_{z}^{2}+m^{2}+2%
\eta +2\tilde{\omega}\left( 2n_{r}+\frac{3}{2}\right) -\frac{A^{2}}{4\tilde{%
\omega}^{2}}\right] ^{1/2}.  \label{A18}
\end{equation}%
and eigen functions 
\begin{equation}
\psi \left( \tilde{r}\right) \sim \tilde{r}^{1/2}\exp \left( -\frac{\tilde{%
\omega}\,\tilde{r}^{2}}{2}\right) L_{n_{r}}^{1/2}\left( \tilde{\omega}%
r^{2}\right) \Longleftrightarrow \psi \left( r\right) \sim \tilde{r}%
^{1/2}\exp \left( -\frac{\tilde{\omega}\,\tilde{r}^{2}}{2}\right)
L_{n_{r}}^{1/2}\left( \tilde{\omega}\,\tilde{r}^{2}\right) .  \label{A19}
\end{equation}%
Next, we wish now to compare this result with the same Schr\"{o}dinger oscillator model reported by Medeirosa and de Mello \cite{R2} (section 4.3), with the correct mapping between the parameters used (i.e., $\tilde{\gamma}_{m}=\left( \tilde{\gamma}^{2}-1/4\right) _{ours}$, $\left\vert m/\alpha \right\vert =\tilde{\gamma}_{ours}$, $\omega =eB_{\circ }/2M=0_{ours}$, $\delta _{L}\sqrt{\Delta }=-A_{ours}$, $\epsilon _{k,m}=\tilde{\lambda}_{ours}$, $0=\eta _{ours}$, $k=\left( k_{z}\right) _{ours}$ and $\Delta =\tilde{\omega}_{ours}$). The comparison between our result in (\ref{A18}) and their result in (49) of \cite{R2} (of course, with the missed terms in their (36), (38), (59), (61), i.e., $k^{2}\rightarrow k^{2}+M^{2}$, and their $k^{2}\rightarrow k^{2}+M^{2}-\delta _{L}^{2}/4\Delta $ in (49) and (51)) indicates that the results are in exact agreement with each other and the truncation order $n$ should be correlated with the radial quantum number $n_{r}$ through the relation $n=2n_{r}$. Moreover, the result in (\ref{A18}) suggests that there is no quantization characterization associated with the parameter $A$ or $\tilde{\omega}$, they are both $(n=2\,n_r)$-independent parameters. This is a brute-force evidence that should be taken into account while dealing with this problem.

A final note on the procedures discussed above is critically unavoidable. For $A=0=B$, the recursion relation (\ref{A11-4}) is safely satisfied but not that of (\ref{11-5}). The relation of (\ref{11-5}) implies that $\gamma^{\prime }=0\Longrightarrow \left\vert \tilde{\gamma}\right\vert =-1/2$ which is neither physically nor mathematically acceptable (similar consequence appear in (9) of Fern\'{a}ndez \cite{R6} where the angular momentum quantum number $\ell $ takes the value $\ell =-1$ as $a=b=0$ in his
relation (4)). Yet in the results reported by Medeirosa and de Mello \cite{R2} (section 4.3), we notice that things are more tragic in the sense that if one sets $\eta _{L}=0$ in their (40), then their $\tilde{\omega}_{1,m}$ of their (52) takes the value $\tilde{\omega}_{1,m}=0$. As a result, their reported energy spectrum collapses into that of free particle energy $E_{k,m,n}=\pm \sqrt{k^{2}+M^{2}}$ (although they still have the harmonic oscillator term but their solution tragically fails for such parametric settings).  This is also reflected on their general solution (their section 4.4). The same happens with the results reported by Ver\c{c}in \cite{R7-1} in equation (22). This should lead us to one conclusion. The above mentioned methodical procedure is insecure/unsafe and its results are unreliable. One has, therefore, to resort to a more reliable methodical proposal like the one very recently discussed in \cite{R8}.

\bigskip \textbf{Data Availability Statement} Authors can confirm that all relevant data are included in the article and/or its supplementary information files.

\end{document}